\documentclass[aps,twocolumn,nofootinbib,tightenlines,superscriptaddress,amsmath,amssymb,final,letterpaper]{revtex4-1}
\usepackage[utf8]{inputenc}
\usepackage{graphicx}
\usepackage{amsmath,amssymb,amsthm}
\usepackage{natbib}
\usepackage{bbold}
\usepackage{calc}
\usepackage{bm}
\usepackage{color}
\usepackage{hyperref}
\usepackage[dvipsnames]{xcolor}

\usepackage{ragged2e} 

\hypersetup{unicode=false, pdftoolbar=true, pdfmenubar=true, pdffitwindow=false, pdfstartview={FitH}, pdfnewwindow=true, colorlinks=true, linkcolor=blue, citecolor=blue, filecolor=magenta, urlcolor=blue}

\usepackage{algorithm}
\usepackage{algpseudocode}

\graphicspath{{./figs/}} 

\newcommand{\except}{\backslash}

\begin{abstract}
Spreading models capture key dynamics on networks, such as cascading failures in economic systems, (mis)information diffusion, and pathogen transmission.
Here, we focus on design intervention problems---for example, designing optimal vaccination rollouts or wastewater surveillance systems---which can be solved by comparing outcomes under various counterfactuals.
A leading approach to computing these outcomes is message passing, which allows for the rapid and direct computation of the marginal probabilities for each node.
However, despite its efficiency, classical message passing tends to overestimate outbreak sizes on real-world networks, leading to incorrect predictions and, thus, interventions.
Here, we improve these estimates by using the neighborhood message passing (NMP) framework for the epidemiological calculations.
We evaluate the quality of the improved algorithm and demonstrate how it can be used to test possible solutions to three intervention design problems: influence maximization, optimal vaccination, and sentinel surveillance.
\end{abstract}

\begin{document}

\title{Message passing for epidemiological interventions on networks with loops}
\author{Erik Weis}
\email{weis.er@northeastern.edu}
\affiliation{Vermont Complex Systems Institute, University of Vermont, Burlington, VT 05405, USA}
\affiliation{Network Science Institute, Northeastern University, Boston, MA, 02115, USA}

\author{Laurent H\'ebert-Dufresne}
\affiliation{Vermont Complex Systems Institute, University of Vermont, Burlington, VT 05405, USA}
\affiliation{Department of Computer Science, University of Vermont, Burlington, VT 05405, USA}
\affiliation{Santa Fe Institute, Santa Fe, NM 87501, USA}

\author{Jean-Gabriel Young}
\affiliation{Vermont Complex Systems Institute, University of Vermont, Burlington, VT 05405, USA}
\affiliation{Department of Mathematics and Statistics, University of Vermont, Burlington, VT 05405, USA}
\maketitle

\section{Introduction}
How can we strategically intervene on a few individual nodes to obtain a desired effect on a network? 
This is the core challenge of the network intervention problem.
It arises in various fields including epidemiology~\cite{chami2017SocialNetwork, nishi2020NetworkInterventions}, organizational and behavioral science \cite{valente2012NetworkInterventions}, social media analysis~\cite{aral2018SocialInfluence}, economics~\cite{banerjee2013DiffusionMicrofinance}, marketing~\cite{van2007ProductDiffusion}, and counterterrorism~\cite{diviak2022DynamicsDisruption}.
And the relevant interventions take various forms. 
Influence maximization~\cite{kempeMaximizingSpreadInfluence2003,pei2020influencer}, for example, seeks to optimize the reach of a campaign through the strategic selection of initial adopters~\cite{domingosMiningNetworkValue2001}.
Targeted immunization~\cite{hebert2013global} aims to contain disease spread by vaccinating key individuals. 
Sentinel surveillance focuses on positioning monitoring sites to detect contagions as quickly as possible~\cite{holmeThreeFacesNode2017}. 

Since interesting network dynamics nearly always incorporate randomness, evaluating intervention quality requires averaging over outcomes---a computationally expensive proposition when searching for the optimal intervention among a large set of possibilities.
This problem is amplified when the intervention space is large or when dynamical outcomes vary widely from realization to realization~\cite{allen2022predicting}.
These computational challenges have inspired a wide range of analytical techniques to compute the distribution of dynamical outcomes, including probability generating functions \cite{allen2022predicting}, approximate master equations \cite{hebert2010propagation,marceau2010adaptive, stonge2022influential}, and message passing \cite{shresthaMessagepassingApproachRecurrentstate2015}.
In this paper, we will focus on the latter as a prospective tool for evaluating the quality of interventions for spreading processes on networks.

This application of message passing is enabled by recent algorithmic advances.
Message passing computes the marginal probabilities of interacting random variables---epidemic state, labels, and so on---using their relationships between neighboring variables~\cite{pearlReverendBayesInference1982,mezardInformationPhysicsComputation2009a}; it has been applied to numerous network problems beyond epidemiology~\cite{karrerMessagePassingApproach2010,shresthaMessagepassingApproachRecurrentstate2015,lokhovInferringOriginEpidemic2014,newmanMessagePassingMethods2023} and intervention design~\cite{altarelliContainingEpidemicOutbreaks2014,moroneInfluenceMaximizationComplex2015}, including percolation \cite{karrerPercolationSparseNetworks2014}, community detection \cite{decelleInferencePhaseTransitions2011}, and the statistical physics of spin glasses  \cite{mezardInformationPhysicsComputation2009a}.
Nevertheless, these classical message passing algorithms break down on networks that have an abundance of short loops, which, critically, includes social networks~\cite{newman2018networks}, the most common target for network interventions.
Recent neighborhood message passing (NMP) methods have partially corrected this problem by explicitly accounting for short loops \cite{cantwellMessagePassingNetworks2019,kirkleyBeliefPropagationNetworks2021,cantwell2023heterogeneous}.
We make use of these methods to study intervention design in social networks with message passing.

Our contributions are threefold.
First, we fuse the dynamic message passing framework described in Refs.~\cite{karrerMessagePassingApproach2010,shresthaMessagepassingApproachRecurrentstate2015} with recent developments in ``loopy'' message passing~\cite{kirkleyBeliefPropagationNetworks2021,cantwellMessagePassingNetworks2019}.
This allows us to track the temporal evolution of the marginal probability of infection under cascade dynamics~\cite{kempeMaximizingSpreadInfluence2003} on networks with loops.
The resulting algorithm can further account for probabilistic and temporal interventions at the level of specific nodes.
Second, we show how this framework can be used to test intervention strategies for discrete-time epidemiological dynamics.
And third, we evaluate the quality of this framework on a real-world network, highlighting where it fails and where it makes accurate predictions.

\section{Message passing for spreading dynamics}
\label{sec:message_passing_for_spreading}
\subsection{Cascade dynamics}
Network intervention design problems rely on \emph{models} of dynamics because they amount to predicting how a complex system will behave under different simulated scenarios. 
For epidemiological interventions, we use the independent cascade model~\cite{kempeMaximizingSpreadInfluence2003}, which describes both social contagion~\cite{aral2018social} and classical epidemic dynamics like the discrete-time Susceptible-Infected~\cite{newmanSpreadEpidemicDisease2002} and Susceptible-Infected-Recovered (SIR)~\cite{kenahSecondLookSpread2007b} models as special cases.
This model assigns two states to nodes: infected (active) or susceptible (inactive).
A cascade is seeded with a set of infected nodes, and an infected node $i$ gets a single chance to infect each of its susceptible neighbors $j\in \partial i$ with probability $p_{ij}$.
The contagion stops when no new infections are possible.

Mathematically, the independent cascade model is parametrized by a network structure, a set of initial infection probabilities $\{s_i\}_{i=1..N}$, and a sparse $N\times N$ matrix $\bm{P}$ of infection probabilities $p_{ij}$.
We encode the state of node $i$ at time $t$ as a random variable $X_i^{(t)} \in \{0,1\}$, equal to $1$ when node $i$ has been infected by time $t$.
The transition probabilities $\Pr(X^{(t)}_i \mid X_i^{(t-1)})$ describing the dynamics are given by:
\begin{subequations}
    \label{eqs:cascade_transitions}
    \begin{align}
        \Pr\bigl(X^{(t)}_i \!= 1\mid X_i^{(t-1)} \!= 0\bigr) &= 1 - \!\prod\limits_{j\in\partial i} (1 - p_{ij})^{
            I_j^{(t-1)}
        }\\
        \Pr\bigl(X^{(t)}_i = 1\mid X_i^{(t-1)} = 1\bigr) 
        &=
        1
    ,\end{align}
\end{subequations}
where $\partial i$ is the set of all the neighbors of $i$ and the indicator $I_j^{(t)} = X_{j}^{(t)} \big[ 1 - X_{j}^{(t-1)}\big]$ determines whether node $j$ can attempt to infect its neighbors at time $t$.

\subsection{Classical message passing solution}
\label{sec:classical_mp}
In the context of intervention design, we are interested in the marginal probability of infection of node $i$
\begin{equation}
    \label{eq:temporal_marginal}
    \pi_i(t) = \sum_{\bm{X}}\mathbb{1}\left[X_i^{(t)}=1\right] P(\bm{X})\
,\end{equation}
where $\mathbb{1}\!\left[s \right]$ is an indicator equal to 1 when $s$ is true, where $\bm{X}$ is the matrix of all node states at all time steps, and where the sum is over all such states.

There are exponentially many terms in this sum, making direct calculation infeasible.
Furthermore, a Monte Carlo estimate of $\pi_i(t)$ by direct simulation, while inexpensive, remains relatively costly once placed inside an intervention optimization loop.
This motivates message passing as a computational shortcut.
We will first focus on computing these marginal probabilities in the infinite time limit, $\pi_i = \pi_i(t \rightarrow \infty)$, but we will return to the temporal marginals $\pi_i(t)$ in Section~\ref{subsec:dmp}.

The starting point of our derivation is the secondary quantity $\pi_{i\except j}$, which is defined as the probability that node $i$ has been infected by a node other than $j$ by time $t\rightarrow \infty$, where $i$ and $j$ are implicitly understood to be neighbors.
This quantity can be calculated by noticing the two ways a node $i$ ultimately gets infected: Either it was initially infected, with probability $s_i$, or it was infected by one of its neighbors.
Excluding $j$ from the second class of events to remain consistent with the definition of $\pi_{i\except j}$ yields
\begin{subequations}
\label{eqs:classical_mp}
\begin{equation}
    \pi_{i\except j} 
    = 
    s_i + (1-s_i) 
    \left[
    1 - \prod_{k \in \partial i \except j} 1 - p_{ki} \pi_{k \except i}
    \right]
\label{eq:message_passing_tree}
,\end{equation}
where $\partial i \except j$ denotes the set of neighbors of $i$ excluding node $j$.
We can then use this auxiliary quantity to compute the node marginals, the quantities of primary interest, as
\begin{equation}
    \pi_{i}
    = 
    s_i + (1-s_i) 
    \left[
    1 - \prod_{k \in \partial i} 1 - p_{ki} \pi_{k \except i}
    \right]
\label{eq:message_passing_tree_marginals}
.\end{equation}
\end{subequations}

This derivation implicitly assumes independence: The probability that no neighbor of node $i$ has infected it is calculated as a product of probability mass functions.
On tree networks, these random variables truly are independent, and Eq.~\eqref{eq:message_passing_tree_marginals} is exact---in the sense that $\pi_i=\lim\limits_ {t\to\infty}\mathbb{E}[X_i^{(t)}]$ where the expectation is taken over all possible cascades~\cite{newmanMessagePassingMethods2023}.
But on networks with many short loops (e.g., triangles), $\pi_{i\except j}$ feeds back on itself through cycles, and the derivation no longer formally holds. 

Nevertheless, the lack of strict independence turns out not to be a big problem when the networks are tree\textit{-like}~\cite{melnikUnreasonableEffectivenessTreebased2011}, and loops are few and long~\cite{allardAccuracyMessagepassingApproaches2019}.
The importance of the message a node passes to itself through a loop of length $\ell$ decays as $O(\langle p\rangle^\ell)$, where $\langle p\rangle$ is the average infection probability.
Thus, long loops have vanishing importance in tree-like networks.
That said, the predictions of the message passing can be significantly in error when networks have many short loops---which is, unfortunately, the case for many of the systems where intervention design is of interest.

The neighborhood message passing (NMP) framework~\cite{cantwellMessagePassingNetworks2019,kirkleyBeliefPropagationNetworks2021,cantwell2023heterogeneous}, which we now describe briefly, corrects for this issue.
We refer the reader to the appendix and Refs.~\cite{cantwellMessagePassingNetworks2019,kirkleyBeliefPropagationNetworks2021} for more details.

\subsection{Neighborhood message passing for cascades}

\begin{figure*}
    \centering
    \includegraphics[width=0.9\textwidth]{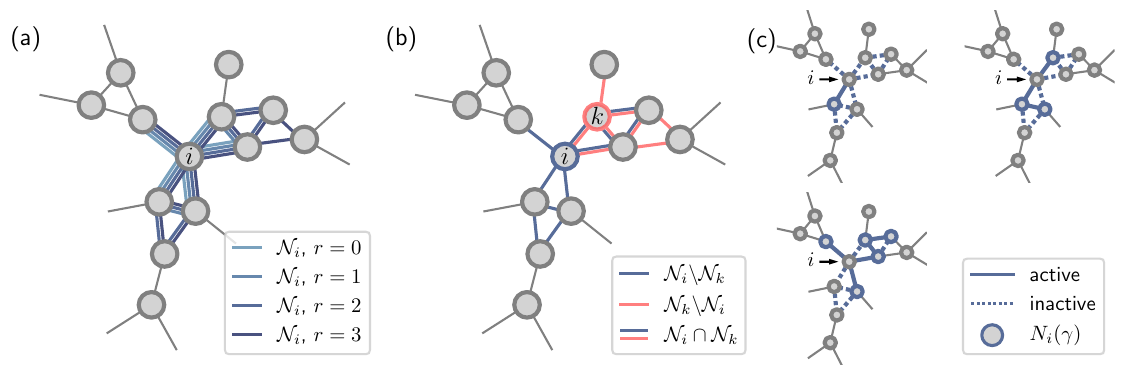}
    \caption{
        \textbf{Neighborhood message passing in complex networks.}~\cite{cantwellMessagePassingNetworks2019,kirkleyBeliefPropagationNetworks2021}
        \textbf{(a)} Construction of nested neighborhoods for different values of $r$. 
        Classical message passing is recovered by setting $r=0$, because nodes in the subgraph induced by $\mathcal{N}_i$ are the neighbors of node $i$, $\partial i$.
        \textbf{(b)} Overlapping neighborhoods, here generated with $r=2$, for two nodes, $i$ and $k$.
        \textbf{(c)} Three example outcomes of the random variable $\Gamma_i$ for the neighborhood $\mathcal{N}_i$ when $r=2$. Each edge in $\mathcal{N}_i$ is shown as capable of spreading the disease (active) or incapable of spreading the disease (inactive). The marginal probability of infection for node $i$, $\pi_i$, is calculated by taking its expected value over all possible configurations $\Gamma_i$.
    }
    \label{fig:conceptual}
\end{figure*}

Neighborhood message passing replaces $\pi_{k\except i}$ with a quantity that characterizes independent random variables under weaker structural assumptions.  
That new quantity is $\pi_{k \except \mathcal{N}_i}$, which is defined as the probability that node $k$ has been infected via a path that does not include the edges in a local \textit{neighborhood} $\mathcal{N}_i$, a small region surrounding node $i$.

Reference~\cite{cantwellMessagePassingNetworks2019} constructs the neighborhood $\mathcal{N}_i$---a set of edges---as follows (see Figure \ref{fig:conceptual}a).
First, we define a primitive cycle of length $\ell$ as a cycle that starts and ends at node $i$ and contains at least one edge not in any shorter primitive cycle.
Then, we define the neighborhood of node $i$ as the edges incident on $i$ and the union of all the edges traversed by any primitive cycle of length $\ell \leq r + 2$, where $r$ represents the size of the neighborhood constructed around $i$.
When $r=0$, NMP reduces to the tree-like approximation of classical message passing since the only primitive cycles of length $\ell=2$ are those that move to a neighboring node and immediately return.
Larger values of $r$ create larger neighborhoods with more edges (and possibly more nodes).
For example, $r=1$ allows for triangles, meaning that $\mathcal{N}_i$ is the edges of the subgraph induced by node $i$ and its immediate neighbors $\partial i$. 

Like classical message passing, the NMP algorithm finds marginal probabilities $\pi_{i}$ using self-consistent equations. 
However, while classical message passing (Sec.~\ref{sec:classical_mp}) computes $\pi_i$ directly from messages $\pi_{i\except j}$, NMP requires a more sophisticated approach.
The marginal probability now depends on the probability $\pi_{k \except \mathcal{N}_i}$ that a node $k$ in the neighborhood of $i$ is infected by an outside source \emph{and} the probability that the cascade reaches node $i$ once it has entered its neighborhood $\mathcal{N}_i$.
To calculate the probability of this sequence of events, we invert the process and first imagine obtaining a set of \emph{active edges} $\gamma$ that \emph{could} have led to an infection, had one of the nodes adjacent to these edges been infected.
Only then do we reason about the probability that node $i$ is reached by a cascade through the edges of $\gamma$.

A random realization $\gamma$ of $\Gamma_i$ is obtained with probability 
\begin{equation}
    \Pr(\Gamma_i=\gamma)=\prod_{(j,k) \in \mathcal{N}_i} p_{jk}^{a_{jk}}(1-p_{jk})^{1-a_{jk}}
\end{equation} 
where $a_{jk}=1$ when nodes $j$ and $k$ are active in configuration $\gamma$, and $0$ otherwise.
(We can think of $\gamma$ as a realization of an inhomogeneous bond percolation process~\cite{allardGeneralExact2015,allard2009heterogeneous} on the subgraph induced by $\mathcal{N}_i$.)

For a given active edge configuration $\gamma$, we define $N_i(\gamma)$ as the set of nodes reachable from $i$.
The conditional infection probability of node $i$ is then the probability that at least one node in $N_i(\gamma)$ is infected
\begin{multline}
    \Pr(X_i = 1 \mid \Gamma_{i} = \gamma)
     \\=s_i + (1 - s_i)
    \left[
    1 - \prod_{k \in N_i(\gamma)} 1 - \pi_{k \except \mathcal{N}_i}
    \right]
    \label{eq:NMP_conditional_marginal}
\end{multline}
This equation again assumes that the various nodes $k$ are infected by a node outside of $\mathcal{N}_i$ independently from one another,  but this independence assumption is now much weaker than before.\footnote{It holds strictly when removing $\mathcal{N}_i$ partitions the graphs in $m$ components where $m$ is the number of nodes with at least one neighbor in the graph induced by $\mathcal{N}_i$.}

We finally obtain the marginal probability $\pi_i$ by summing over all possible configurations of $\Gamma_i$ as
\begin{equation}
    \pi_{i} = \sum_{\gamma} \Pr(X_i =1 \mid \Gamma_{i} = \gamma) \Pr(\Gamma_i=\gamma)
    \label{eq:NMP_marginal}
.\end{equation}

Self-consistent equations for the messages are required to close the system of equations.
Much like in classical message passing, only small modifications to the marginals' calculation are needed to obtain these equations.
Whereas we previously removed $j$ from the product in Eq.~\eqref{eq:message_passing_tree_marginals} to obtain Eq.~\eqref{eq:message_passing_tree}, we now need to remove all \textit{edges} in $\mathcal{N}_j$ from the neighborhood of $\mathcal{N}_i$ before calculating the marginal.
This change amounts to defining a new random variable $\Gamma_{\mathcal{N}_i \except \mathcal{N}_j}$, which corresponds to the outcome of bond percolation on the edge subset $\mathcal{N}_i  \except \mathcal{N}_j$ (see Fig.~\ref{fig:conceptual}b).
Finally, we obtain the marginal probability by summing over the outcomes of $\Gamma_{\mathcal{N}_i \except \mathcal{N}_j}$ to obtain
\begin{equation}
    \pi_{i\except \mathcal{N}_j} = \sum_{\gamma} \Pr(X_i =1 \mid \Gamma_{\mathcal{N}_i \except \mathcal{N}_j} = \gamma) \Pr( \Gamma_{\mathcal{N}_i \except \mathcal{N}_j}=\gamma)
    \label{eq:NMP}
,\end{equation}
where the update equation for $\Pr(X_i =1 \mid \Gamma_{\mathcal{N}_i \except \mathcal{N}_j} = \gamma)$ is analogous to Eq.~\eqref{eq:NMP_conditional_marginal} with the only difference being that $\gamma$ is a realization of $\Gamma_{\mathcal{N}_i \except \mathcal{N}_j}$, i.e., a configuration of the conditional neighborhood $\mathcal{N}_i \except \mathcal{N}_j$ (constructed with the procedure shown in Fig.~\ref{fig:conceptual}b).

\subsection{Dynamic neighborhood message passing}
\label{subsec:dmp}
Recall that our goal is to evaluate network interventions using message passing as a computational shortcut.
Since several interventions rely on the timing of infection events---e.g., interventions designed to monitor early outbreaks---we now extend the NMP framework to track the temporal dynamics of the cascade.

As was shown in Ref.~\cite{lokhovInferringOriginEpidemic2014}, using the initial conditions of a cascade as the initial state of the system of (classical) message passing equations allows one to interpret each step of the algorithm as the steps of a discrete-time SIR process. 
This leads to the following definitions of dynamical messages
\begin{align*}
    \pi_{i\except j}(t+1)
    &= 
    s_i + (1-s_i) 
    \left[
    1 - \prod_{k \in \partial i \except j} 1 - p_{ki} \pi_{k \except i}(t)
    \right],
\end{align*}
where the messages at time $t+1$ are defined in terms of the messages at time $t$ under a synchronous update schedule.
(With similar modifications for the marginal calculation.)

This idea can be translated to NMP almost directly, though a critical complication arises: We now need to account for all the possible pathways through which $i$ was infected within its neighborhood, all with potentially different impacts on the timing of the infection depending on their length.
To demonstrate this idea, we explain how to calculate the marginal probability $\pi_i(t)$ though, again, similar adjustments are needed for calculating the messages $\pi_{k\except \mathcal{N}_i}(t)$.

For each node $k$ reachable from node $i$, we let $\ell_k$ be the length of the shortest path from node $i$ to $k$ using only the active edges of $\gamma$, a realization of $\Gamma_i$.
(We do not index $\ell_k$ to simplify the notation, but one should remember that this quantity is defined with respect to node $i$.) 
If $k$ is infected at time $t-\ell_k$, then $i$ will be infected at time $t$.
Thus, the conditional probability of infection for node $i$ at time $t$ is the probability that at least one node in $N_i(\gamma)$ is infected \emph{early enough} for the cascade to reach $i$ by time $t$:
\begin{multline}
    \Pr\left(X_i^{(t)}=1\ \Big|\ \Gamma_i=\gamma\right)\\ = 
    s_i + (1 - s_i) \prod_{k \in N_i(\gamma)}
    \left[
    1 - \pi_{k\except \mathcal{N}_i} (t-\ell_k)
    \right]
    \label{eq:dynamicNMP_core}
.\end{multline}
While this change is minor, it does translate to additional bookkeeping.
For each outcome $\Gamma_i$ (or $\Gamma_{\mathcal{N}_i \except \mathcal{N}_j}$), we must store the values of $\pi_{i\except j}$ for the past $\ell_{\max}$ steps of the algorithm, where $\ell_{\max}$ is the longest shortest path from any node $i$ to a node in its neighborhood.\footnote{In practice, we have found that the marginal probabilities $\{\pi_i(t)\}$ all reach a stable value after only a few time steps $t$ (on the order of the network's diameter), so our reference implementation stores the entire history instead of a variable number of messages for each neighborhood.}

\subsection{Implementation details}
We provide a Python implementation of the dynamic NMP algorithm, available online~\cite{zenodo}.
In addition to the techniques described above, we incorporate a few optimizations.

\subsubsection{Neighborhood Monte Carlo}
Practically speaking, we should be concerned about the number of possible outcomes of the random variables $\Gamma_i$ and $\Gamma_{\mathcal{N}_i \except \mathcal{N}_j}$.
Even when we set $r=0$, the sums in Eqs.~\eqref{eq:NMP_marginal} and \eqref{eq:NMP} contain $2^q$ terms, where $q$ is the degree of node $i$. 
Exact NMP calculations are thus prohibitively costly for nearly all networks.

The standard solution is to replace the sum in Eqs.~\eqref{eq:NMP_marginal} and \eqref{eq:NMP} with a sampling approximation, for example,
\begin{equation}
    \pi_{i\except \mathcal{N}_t}(t) \approx \sum_{m=1}^M \Pr\left(X_i^{(t)} = 1\ \Big|\ \Gamma_{\mathcal{N}_i \except \mathcal{N}_j} = \gamma_m\right)
    \label{eq:sampling_approx}
,\end{equation}
where $\gamma_m$ is a Monte Carlo sample of the configuration $\Gamma_{\mathcal{N}_i \except \mathcal{N}_j}$, and $M$ is the number of samples.
This is roughly the solution we favor, though we use additional computational tricks specific to cascade dynamics~\cite{cantwellMessagePassingNetworks2019}---see Appendix~\ref{appendix:sampling} for details.
The samples are generated once and used throughout the algorithm's execution to facilitate convergence.

\subsubsection{Hyperparameters}
Setting the size of the neighborhood $r$ and the number of samples $M$ is another important computational consideration, as these hyperparameters control the complexity and precision of the resulting NMP algorithm.
For $r=0$, the algorithm simplifies to classical pairwise message passing and is thus cheap but typically inaccurate.
The accuracy increases for large values of $r$, but this may also force us to perform Monte Carlo simulations on very large neighborhoods that can easily span the entire network if $r$ is too large.
The resulting algorithm will be as accurate as simple Monte Carlo simulations of the whole cascade dynamics but also far too slow, as we will have to perform these simulations in the neighborhood of every node, deduplicating the costs needlessly.

In practice, we tested various values of $r$ and investigated the impact of this hyperparameter on intervention accuracy and computational costs.
We found that low values of $r$ are often sufficient, c.f. Sec.~\ref{sec:results}.
As for the number of samples $M$, efficient sampling algorithms exist \cite{newmanFastMonteCarlo2001,weis2024robust}, so its impact on the overall complexity is less pronounced (a simple multiplicative factor).
In practice, we found that the sampling accuracy is quite good after surprisingly few samples $M$, regardless of the value of $r$ (see Appendix~\ref{appendix:costs} for details), so we will focus chiefly on the effect of $r$ in the following sections.

\begin{figure*}
    \centering
    \includegraphics[width=0.9\textwidth]{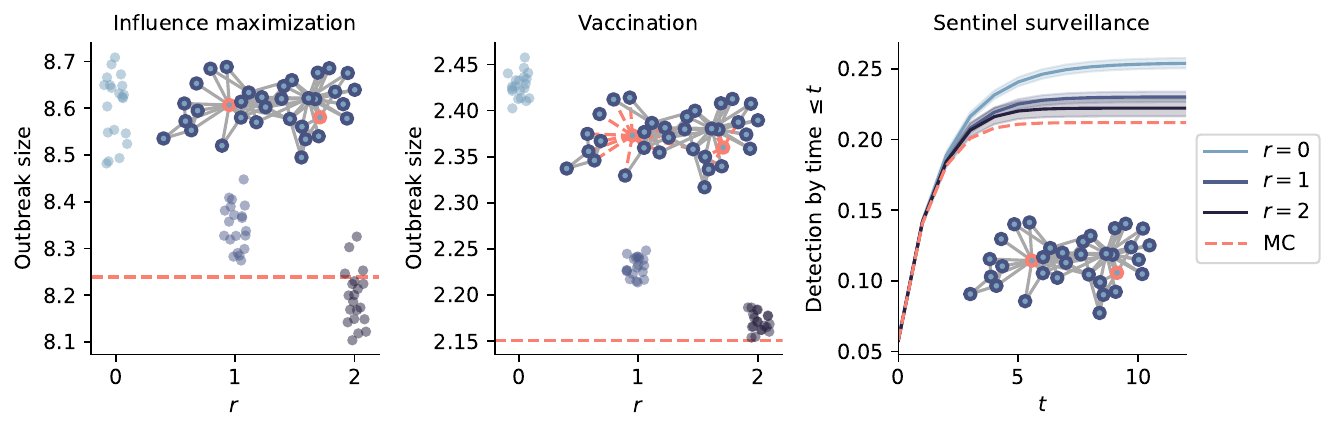}
    \caption{
        \textbf{Quality of intervention as evaluated by neighborhood message passing.}
        We calculate the estimated impact of three interventions with Monte Carlo simulations (MC, dashed lines) and the neighborhood message passing algorithm (NMP, markers and solid lines) on the karate club graph~\cite{zachary1977_InformationFlowModel}, shown in the insets.
        Nodes that receive an intervention are shown in red, and a uniform infection probability of $p=0.15$ is used on all edges. 
        We test three neighborhood sizes, $r\in\{0,1,2\}$. 
        For all problems, we use $M=1,500$ samples for each neighborhood and show results for $20$ different realizations of the dynamic NMP algorithm (Eq. \ref{eq:message_passing_sentinel}). 
        For influence maximization and vaccination, we show the expected outbreak size for each realization.
        For sentinel surveillance, we show the evolution of the cumulative detection probability as a function of $t$.
    }
    \label{fig:conceptual_IP}
\end{figure*}

\section{Interventions in the neighborhood message passing framework}
\label{sec:intervention}

With the dynamic NMP framework in place, we now turn to intervention design.
Recall that network intervention design, broadly construed, is the strategic manipulation of a network's nodes to achieve a desired dynamical outcome.
The epidemiological interventions we study here include targeted vaccination and sentinel surveillance \cite{holmeThreeFacesNode2017}, as well as various interventions that involve quenched initial conditions, such as influence maximization \cite{kempeMaximizingSpreadInfluence2003} and adversarial versions thereof designed for defense against worse case outbreaks \cite{khim2019adversarial}; see Fig.~\ref{fig:conceptual_IP} for an overview.

\subsection{Quenched dynamics: influence maximization}
The influence maximization problem seeks a small subset of seed nodes that maximizes the expected size of outbreaks~\cite{kempeMaximizingSpreadInfluence2003}.
Thus, we must be able to predict the size of an outbreak when specific nodes are chosen as the seeds.

We made the initial conditions explicit parameters of the dynamic NMP framework, which means that testing various seed sets is extremely straightforward: We can start the cascade at a specific set of nodes by specifying the vector of probabilities $\bm{s}$ accordingly.
For instance, if we want to initialize a cascade in which node $i$ is the only infected node, we simply have to set $s_i=1$ and $s_j=0$ for all other nodes $j$.
This lets us study influence maximization, robustness, and defense against targeted infection.

To design an influence maximization \emph{intervention}, we define a function $Q_{\mathrm{I}}(S)$ that captures the quality of set $S$ as its expected total outbreak size
\begin{equation}
    Q_{\mathrm{I}}(S) = \sum_i \pi_i
,\end{equation}
where it is understood that the marginals are calculated by setting $s_i=1$ for all nodes in $i\in S$ and $s_i=0$ otherwise.
The first panel of Fig.~\ref{fig:conceptual_IP} shows that the NMP estimate of $Q_I(S)$ rapidly converges to the Monte Carlo estimate as the neighborhood size $r$ increases.
This function can then be used as a maximization objective over fixed-sized subsets to optimize spread (or characterize the worst-case scenario).

\subsection{Vaccination}
Next, focusing on vaccination, we handle immunity by altering the probability $p_{ij}$ that node $i$ will infect node $j$ if it is infected.
This strategy is quite general and can describe complex vaccination outcomes, including partial immunity.
For instance, a fully immunized node $i$ cannot spread a cascade or become infected, and this can be expressed by setting $p_{ij}=0$ for all nodes $j$ in its neighborhood $\partial_i$, as well as setting $p_{ki} = 0$ for any edge including $i$ in all other neighborhoods.
This second condition ensures infection pathways in neighborhoods do not pass through immunized nodes.
The independent cascade model can also describe partial vaccine efficacy, which could happen via $i$'s reduced transmissibility, $j$'s reduced susceptibility, or a combination of both \cite{gozzi2023estimating}. 
In any of these cases, partial immunity can be represented as changes to the infection probabilities $p_{ij}$.
For the present paper, however, we will focus on complete immunity and analyze interventions in which a certain subset of nodes completely stops spread.

Once a vaccination schedule has been chosen, we calculate its quality function with a function reminiscent of influence maximization,
\begin{equation}
    Q_{\mathrm{V}}(S) = - \sum_i \pi_i,
\end{equation}
the negative total expected outbreak size when nodes $i\in S$ are vaccinated.
Again, one can treat $Q_{\mathrm{V}}(S)$ as a maximization objective over subsets of a fixed size (say, if the vaccine budget is limited), possible rollout strategies, partial immunization levels, etc.
Alternatively, and like we do here, $Q_{\mathrm{V}}(S)$ can be used to compare the quality of a few constructed sets and compare planned interventions $S$. 
The second panel of Fig.~\ref{fig:conceptual_IP} shows that the NMP estimate of $Q_{\mathrm{V}}(S)$ approaches the true value as $r$ increases.

\subsection{Sentinel surveillance}
Finally, we can also evaluate the quality of a set of sentinels with dynamic NMP.
Good sentinels are nodes that will pick up on an emerging cascade as rapidly as possible.
Though this problem is not strictly speaking an intervention, since surveillance is passive and does not affect outcomes, it is nonetheless a decision problem that involves network dynamics, and the DMP framework can again help compute outcomes. 

Designing a good sentinel surveillance objective is not trivial.
For example, we could calculate the probability that at least one sentinel in a set $S$ has been infected by time $t$.
If sentinels were independent, this probability would be given by
\begin{equation*}
    1 - \prod_{s\in S} \bigl[1-\pi_s(t)\bigr]
,\end{equation*}
In reality, we expect nearby sentinels to be strongly related; failing to take these dependencies into account could lead one to choose sentinels in the same area of the network (where they are all expected to activate quickly), even though this approach is less than optimal (because coverage is less comprehensive).

A better objective can be constructed by modifying the NMP equations to obtain the probability $\pi^{(S)}_i(t)$ that node $i$ has been infected by a path that does not contain a node in the sentinel set $S$ by time $t$---more on this shortly.
This then allows us to calculate the probability that the cascade makes its way to the sentinel set by time $t$ as
\begin{equation}
     \pi_{S}(t) = 1 - \prod_{i \in S} \left[1 - \pi^{(S)}_i(t)\right].
    \label{eq:sentinel_quality}
\end{equation}
Finally, we define the quality of a set of sentinels as the expected time to detection
\begin{multline}
    Q_{\mathrm{T}}(S) = (1- \pi_S)\cdot D(G) \\+ \pi_{S}
        \sum_{t=0}^\infty t \Big[\pi_S(t) - \pi_S(t-1)\Big]
    \label{eq:quality_sentinel}
,\end{multline}
where  $1 - \pi_{S}$ is the probability that the sentinel set is never infected (given by $\pi_S(t)$ in the limit $t\to\infty$), and $D(G)$ is the diameter of the network.
The first term accounts for the scenario in which the sentinels never detect an existing cascade (in which case the detection time is set to $D(G)$, the upper bound), while the second term is, more straightforwardly, the expected detection time when sentinels discover the cascade. 
(The term $\pi_S(t) - \pi_S(t-1)$ corresponds to the probability that the cascade is detected precisely at time $t$ by the sentinels.)

To complete the calculation, we note that $\pi^{(S)}_i(t)$ can be obtained by modifying the calculation of the marginal probabilities $\pi_i(t)$, as
\begin{multline}
    \Pr\left(X_i^{(t)}=1\ \Big|\ \Gamma_i=\gamma\right)\\ = 
    1 - \prod_{k \in N_i(\gamma)}
    \left[
    1 - \mathbb{1}_{[k \notin S]} \pi_{k\except \mathcal{N}_i} (t-\ell_k)
    \right]
\label{eq:message_passing_sentinel}
.\end{multline}
The indicator prevents sentinels from spreading the cascade further, as this would violate the definition of $\pi^{(S)}_i$.
Furthermore, the variable $\Gamma$ also changes, where we set $p_{ij} = 0$ for all edges adjacent to a sentinel.
The marginal $\pi^{(S)}_i(t)$ can then be obtained as before by sampling configurations $\gamma$.

\section{Results}
\label{sec:results}
Our main question is this: How well does NMP approximate cascade dynamics in the context of intervention design?
It can be operationalized in two different ways.
First, how accurately do NMP objectives approximate those calculated with Monte Carlo simulations of the cascade process? 
Second, how do intervention strategies differ when designed with either of these objectives?

\begin{figure*}
    \centering
    \includegraphics[width=0.9\linewidth]{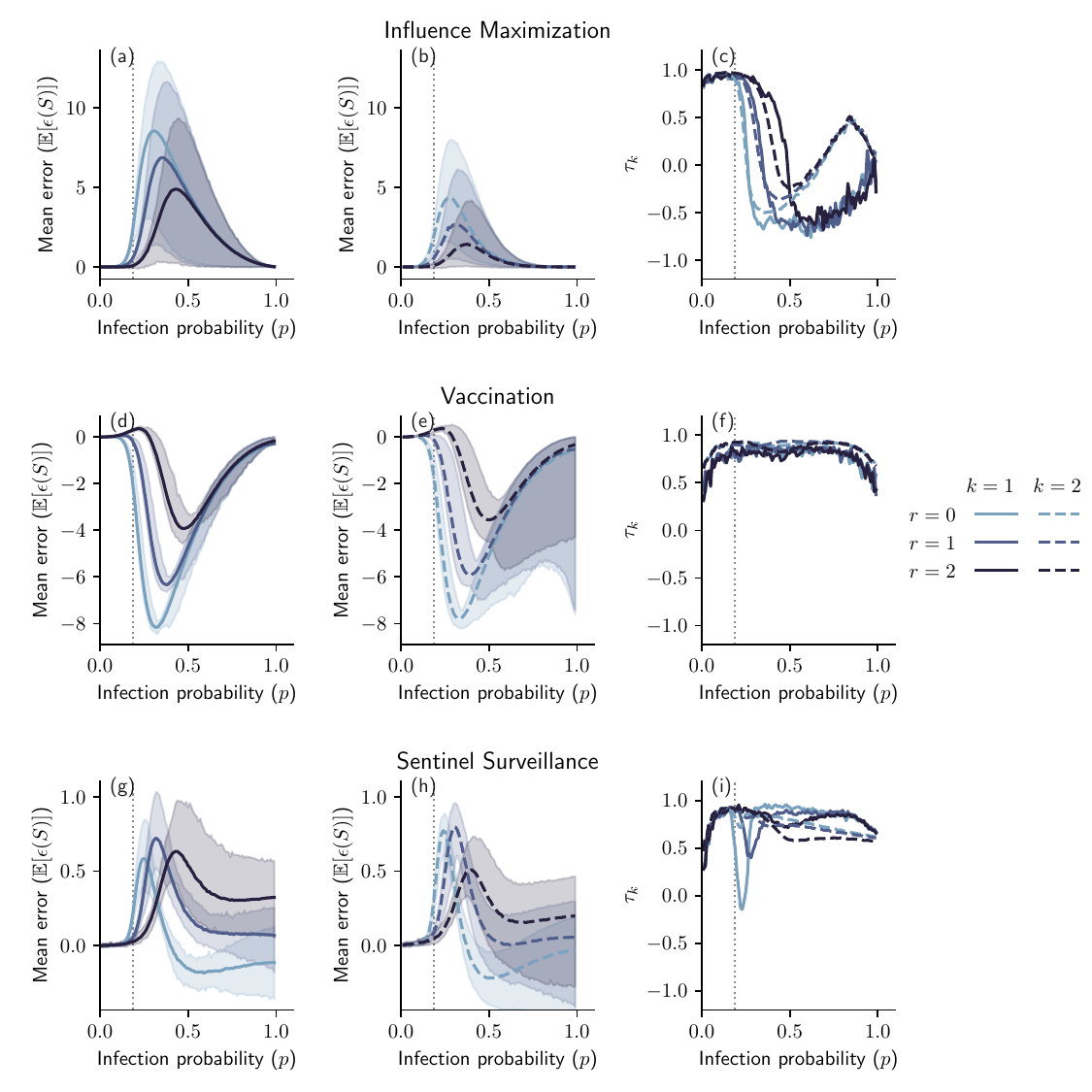}
    \caption{
    \textbf{Performance of NMP on the karate club network~\cite{zachary1977_InformationFlowModel}.}
    \textbf{(a,b)} Difference between the true expected outbreak size and the NMP approximation averaged over all seed sets of size (a) $k=1$ and (b) $k=2$. 
    \textbf{(c)} Kendall rank correlation coefficient of the true set qualities and those predicted by NMP. 
    \textbf{(d-f)} Same as (a-c), except the intervention is now a targeted vaccination. 
    \textbf{(g-i)} Same as (a-c), except the intervention is now surveillance and the error is calculated using the expected time to detection, Eq.~\eqref{eq:quality_sentinel}, as the quality function. 
    Error bars show a 95th percentile interval.
    NMP results were run with $M=1,500$ Monte Carlo samples for all neighborhoods. 
    MC estimates of $Q^*(S)$ were computed with $10^6$ simulations.
    Vertical dotted lines show the critical point of the $r=0$ NMP system without intervention. }
    \label{fig:results}
\end{figure*}

Answering these questions requires extensive simulation: We must evaluate hundreds of intervention sets at various infection probabilities $p$ using both approaches.
The computational costs of such an analysis are hefty, so we will focus on a case study of the karate club network \cite{zachary1977_InformationFlowModel}---a small network with many short loops, where the corrections of NMP are expected to make a sizeable difference. 

\subsection{Neighborhood message passing delays the onset of errors}

For our first analysis, we calculate the absolute difference between expected quality $Q^*(S)$ (computed with Monte Carlo simulation) and the quality estimated by message passing, $Q(S)$, corresponding to
\begin{equation}
    \epsilon(S) = Q(S) - Q^*(S)
.\end{equation}
The results are shown in the first and second columns of Fig.~\ref{fig:results} for all three intervention types---influence maximization (first row), vaccination (second row), and sentinel surveillance (third row).
The curves show the error averaged over all intervention sets of size $k=1$ (first column) and $k=2$ (second column).
This error, $\mathbb{E}\left[\epsilon(S) \right]$, can be understood as an expected value when interventions are sampled uniformly at random from all possible interventions of a fixed size $k$~\cite{holmeThreeFacesNode2017}.

Across all three problems, we observe that the expected error starts growing when the probability of infection reaches the epidemic threshold.\footnote{The network is finite and does not formally have an epidemic threshold. 
Nonetheless, because of the loops in the structure, classical message passing equations can lead to a macroscopic percolating cascade for $p>p_c=0.189$~\cite{karrerPercolationSparseNetworks2014,allardAccuracyMessagepassingApproaches2019}. We refer to this point as the ``epidemic threshold'' even if this is a slight abuse of nomenclature.}
Importantly, using larger neighborhoods $r$ delays the onset of errors later into the critical regime, at larger values of $p$.

The critical regime is where judicious interventions can have the most impact.
Indeed, when the infection probability $p$ is small, most cascades stop rapidly, even if the seed set is chosen strategically.
Conversely, as $p$ increases, the expected outbreak size approaches the size of the network regardless of the intervention made.

We also find that increasing the neighborhood size $r$ consistently reduces the expected error $\epsilon(S)$ for both the influence maximization and vaccination problems.
While this reduction yields helpful improvements in the region around $p_c$, long loops in the network eventually lead message passing to consistently overestimate the expected outbreak size.
In the case of influence maximization, as $p\rightarrow 1$, the error trivially reduces to zero because any infection anywhere in the network leads to total infection.
For vaccination, the error persists even when $p=1$ because of how message passing understands initial conditions.
Our simulations are initialized with a single seed, meaning the initial conditions of individual nodes are correlated, as only one can be the seed; however, message passing assumes independent and uniform initial conditions.
When vaccination breaks the network into two or more, simulations will produce complete infection in just one of the connected components.
Message passing should do the same, but within each component, long loops in the system lead to erroneously predicting the complete infection of both components.

Increasing the neighborhood size $r$ also improves the quality of estimates for sentinel surveillance, but not across all values of $p$. 
For this problem, dynamical message passing generally overestimates the quality of sentinel sets at a rate which varies with $p$ and neighborhood size $r$. 
The source of this error is the same for all parameter regimes: that dynamical message passing overestimates the probability of a sentinel set detecting an outbreak at a late time [see Fig. \ref{fig:coreness_ss} (a)]---the reason being that errors caused by long loops take a longer time to propagate through the system (see Figure \ref{fig:conceptual_IP}).
As we see with the other problems, the global overestimation of the total outbreak size increases with $p$, leading to an overestimation that a sentinel set detects an outbreak at all [see Fig. \ref{fig:coreness_ss} (b)].
Taken separately, the late-time bias and global error are easily understood sources of error.
However, their tradeoff leads to somewhat strange predictions for the absolute quality of sentinels, due to the way they are combined in Eq. \ref{eq:quality_sentinel}.
For low $p$, the error is dominated by the overestimation of detecting the outbreak late in the process, because the global overestimation of the outbreak size is mild. 
For high $p$, the global error dominates, leading to an underestimate of the probability that a sentinel set never detects the outbreak and thus underweighting of the penalty term for this process.
Regardless of the direction of the error, a larger neighborhood size $r$ brings the infection rate closer to zero in the region around the critical threshold.

\subsection{Interventions can be misaligned}
Our second analysis focuses on ranking consistency: Do message passing and simulations prioritize intervention sets similarly when evaluated by dynamic importance objectives?
This matters because errors in the magnitude of the quality function are inconsequential if the relative order of interventions is maintained.

We measure the consistency of the rankings using the Kendall rank correlation coefficient $\tau_k$.
Specifically, we first compute the quality of all interventions of a fixed size $k$ with Monte Carlo simulations and NMP at a specific neighborhood size $r$ and infection probability $p$. 
This yields two total orderings of the interventions.
We then calculate the correlation of the resulting ranking, which gives us a measure of the agreement between the two approximation methods.

Our results are reported in the third column of Fig.~\ref{fig:results}, where we show the rank correlation of all interventions of size $k=1,2$ for the karate club network.

The behavior of the rank correlation differs for each problem, although all of our results are explained by the same mechanism. 
As we have already shown, NMP and Monte Carlo simulations agree broadly below the critical threshold $p_c$.
Thus, we expect that similar interventions will be chosen regardless of the calculation method used in this regime, and Fig.~\ref{fig:results} confirms it.
\footnote{There is no signal when $p=0$, and the correlation correctly goes to $0$ then.}

\begin{figure}[t]
    \centering
    \includegraphics[width=1.0\linewidth]{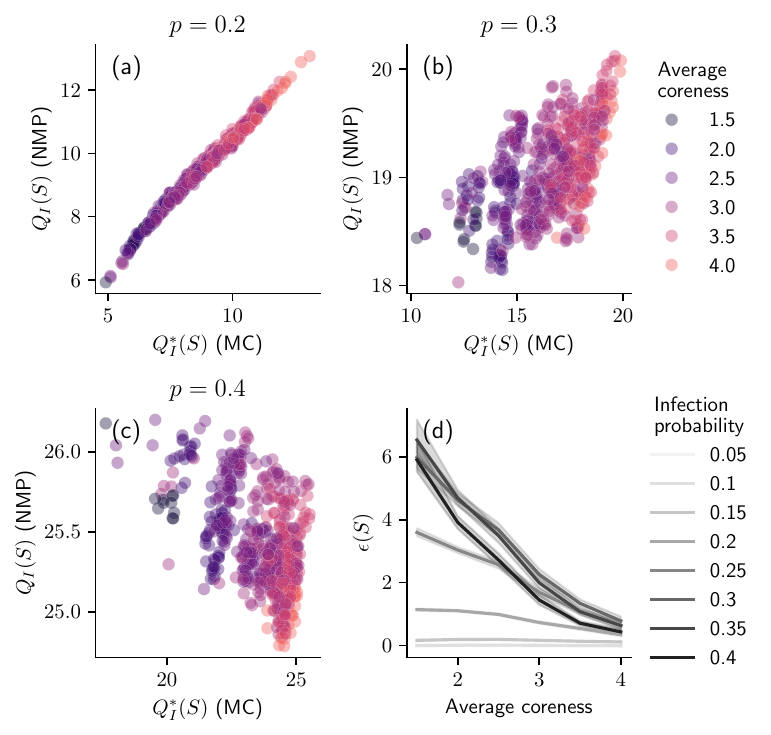}
    \caption{\textbf{Behavior of the influence maximization objective function across seed sets.} 
    \textbf{(a-c)}, we compare the expected final outbreak size $Q_I(S)$ predicted by Monte Carlo and NMP ($r$ = 1).
    Each point corresponds to an outbreak seeded to a different set of size $k=2$. 
    The Monte Carlo averages are calculated with $10^6$ simulations.
    The points are colored by the average coreness~\cite{newman2018networks} of the nodes in the set.
    \textbf{(d)} Difference between true and predicted outbreak sizes as a function of coreness for various infection probabilities.
    }
    \label{fig:source_of_difference}
\end{figure}

The story is different when $p$ is above $p_c$.
Despite the neighborhood corrections, NMP tends to overestimate the probability that an outbreak will reach the whole network, and this is particularly true of cascades seeded on the periphery of the network, see Fig.~\ref{fig:source_of_difference}.
As a result, NMP generally overweights the danger posed by peripheral nodes, something we can quantify through node coreness, the highest $k$-core (or subgraph where all nodes have degree at least $k$) in which a node is found~\cite{newman2018networks}.

For the influence maximization problem, this mis-calibration translates to a precipitous drop in performance when $p$ is above the critical threshold; see Fig.~\ref{fig:results}~(c).
The transition to the error-prone regime is again pushed to higher values of $p$ when $r$ increases.
As $p$ goes well past $p_c$, the rank correlation even switches to an anti-correlated regime. 
This result is explained by Fig.~\ref{fig:source_of_difference}: As outbreaks become larger, the overestimation of outbreaks from nodes of low coreness becomes stronger than the actual difference in outbreaks across seeds.

\begin{figure}
    \centering
    \includegraphics[width=1.0\linewidth]{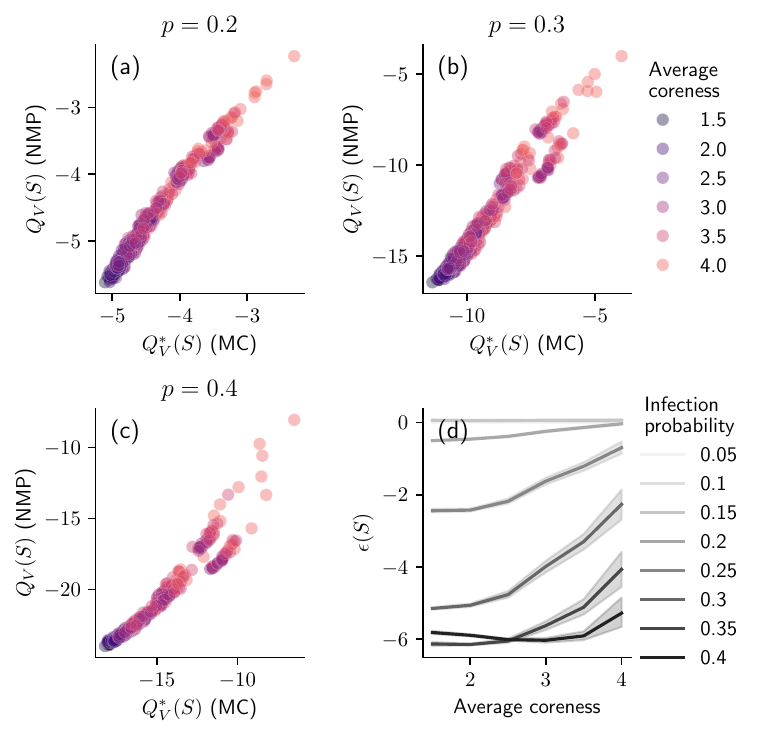}
    \caption{\textbf{Behavior of the vaccination objective function across seed sets.}
    \textbf{(a-c)} The negative expected final outbreak size $Q_V(S)$ predicted by Monte Carlo and NMP.
    Configurations are the same as in Fig. \ref{fig:source_of_difference} {(a-c)}.
    \textbf{(d)} Difference between true and predicted outbreak sizes as a function of coreness for various infection probabilities.
    }
\label{fig:coreness_vaccination}
\end{figure}

For the vaccination problem, the rankings generally agree [see Fig.~\ref{fig:results}~(f)] regardless of the large errors in absolute value [Fig.~\ref{fig:results}~(d,e)].
NMP inoculates central nodes because they lead to large outbreaks if left unattended, and Monte Carlo does the same.
As with influence maximization, the overestimation of outbreaks is negatively correlated with coreness; however, this effect is not strong enough to invert the rankings of sets; see Fig.~\ref{fig:coreness_vaccination}.
One reason for this reduced impact is that vaccination sets are evaluated on outbreaks that are seeded across the network, not exclusively in either the core or periphery.
Unlike influence maximization with peripheral seeds, the core is expected to be most infected for the problem of vaccination, leaving less room for overestimation.
Another reason is that vaccinating the core actively reduces the number of small loops in the network, limiting sources of error.

\begin{figure} 
    \centering
    \includegraphics[width=1.0\linewidth]{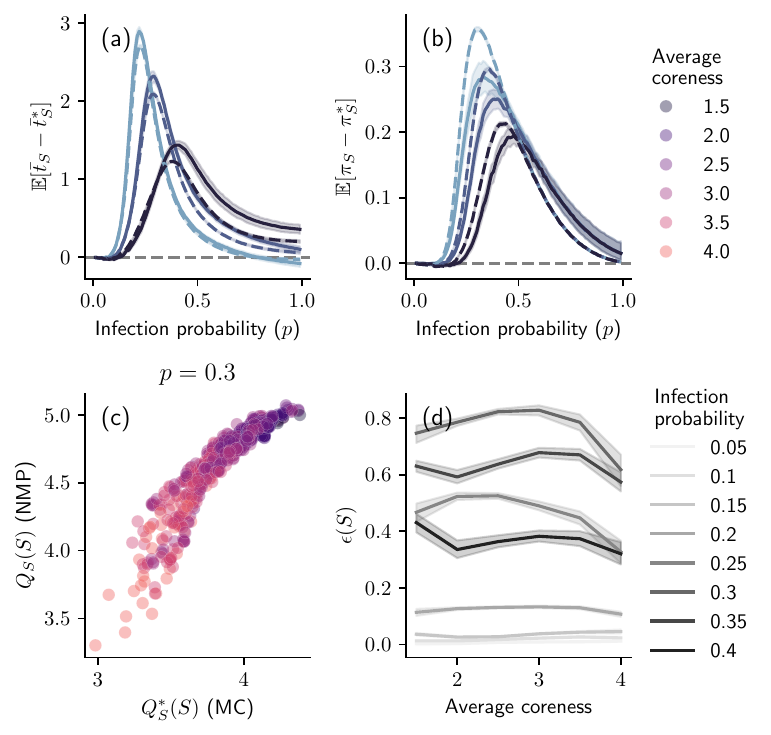}
    \caption{\textbf{Behavior of the sentinel surveillance objective function across seed sets.}
    \textbf{(a)} The difference in average expected infection times of a sentinel node, as predicted by Monte Carlo ($\bar{t}^*_S$) and NMP ($\bar{t}_S$).
    \textbf{(b)} The difference in the probability that a sentinel node gets detected by time $t \rightarrow \infty$, as predicted by Monte Carlo $(\pi_S^*)$ and NMP ($\pi_S$).
    \textbf{(c)} The expected sentinel quality $Q_T(S)$ predicted by Monte Carlo and NMP.
    Configurations are the same as in Fig. \ref{fig:source_of_difference} {(a-c)}.
    \textbf{(d)} Difference between true and predicted outbreak sizes as a function of coreness for various infection probabilities.}
    \label{fig:coreness_ss}
\end{figure}

Finally, for sentinel surveillance, we observe a temporary drop in performance above $p_c$, and it is more pronounced for sets of size $k=1$ than for sets of size $k=2$.
The disagreement arises because NMP prefers more central nodes than Monte Carlo simulations.
NMP predicts larger outbreaks when seeded on the periphery, and these outbreaks are best detected by placing nodes in the center, as a large outbreak will inevitably pass through the center.
Monte Carlo simulations better capture the possibility that these outbreaks will die a stochastic death, and slightly more peripheral nodes are thus better sentinels; see Fig.~\ref{fig:coreness_ss}.

\section{Discussion}

We proposed a message passing method for computing the temporal marginals for the independent cascade model on networks with many short loops.
The method is useful for evaluating the quality of network interventions, as it strictly improves the accuracy of calculations compared to other analytical frameworks.

We found that the main effect of introducing neighborhood corrections is to delay the onset of a regime where large errors occur.
Indeed, message passing starts to err around the critical threshold of an epidemic, and we have found that increasing the neighborhood size pushes this behavior to larger values of the infection probability.
We found that message passing struggles with high infection probabilities despite accounting for loops, which seems inevitable for any approximation method.
Furthermore, in this regime, we also found that message passing seems to show a particular bias towards certain interventions, affecting the overall node rankings, though the bias is not as strong for larger neighborhoods---except in the case of sentinel surveillance, where one is sometimes, paradoxically, better off using less drastic corrections.
Previous work has suggested that this bias is due to the structural property of $k$-coreness \cite{allardAccuracyMessagepassingApproaches2019}, though more work is needed to understand this effect thoroughly.

Recent literature suggests a few ways to further improve the performance of NMP, and it will be interesting to investigate their impact on the design of network interventions.
For example, some methods attempt to improve the tradeoff between accuracy and complexity, e.g., by letting the neighborhood size $r$ vary from node to node or by using mean-field approximations instead of Monte Carlo sampling when the neighborhood is sufficiently large~\cite{cantwell2023heterogeneous}.
Another line of previous work has focused on small motifs, such as triangles \cite{radicchiLocallyTreelikeApproximation2016} or fully connected cliques \cite{yoonBeliefpropagation2011}, that may appear in a network.
The rough idea is to pre-calculate dynamical outcomes analytically, offering a potential performance improvement~\cite{allardGeneralExact2015,laurence2018exact}.
By definition, any node that is attached to a clique will contain all the edges of that clique in its neighborhood, assuming $r>0$.
Thus, accounting for cliques would not necessarily improve the estimation of NMP, but it may allow us to compute the probability $P(X_i = 1 | \Gamma_i)$ more efficiently.
Relatedly, the idea of finding highly connected network regions also suggests further work on hybrid message passing techniques.
For example, there may be general ways of dividing the network into quasi-independent regions and nesting message passing algorithms within each other.
Some work of this flavor has already been explored for exact percolation algorithms on small networks \cite{allardGeneralExact2015}.

Finally, we also note that NMP could be leveraged more directly by network intervention design algorithms.
For instance, instead of testing all sets traversing the space of intervention with greedy submodular search~\cite{kempeMaximizingSpreadInfluence2003}, one might use physics-inspired maximization methods~\cite{altarelliContainingEpidemicOutbreaks2014} with NMP corrections, or improve the accuracy of gradient-based methods~\cite{lokhov2017optimal}.
These issues, however, we leave for future work.

\section*{Acknowledgement}

Research reported in this publication was supported by the National Institute of General Medical Sciences (NIGMS) of the National Institutes of Health under award number P20GM125498.
We are grateful for support from the Vermont Complex Systems Institute. Computations were performed, in part, on the Vermont Advanced Computing Center.
The authors thank Alec Kirkley and Bren Case for helpful conversations.

\appendix

\section{Sampling neighborhood outcomes}
\label{appendix:sampling}

As stated in the main text, the random variables $\Gamma_i$ can be understood as an inhomogeneous bond percolation process taking place in the neighborhood of each node. 
This process yields samples ${ \gamma_1, \gamma_2, \ldots, \gamma_M }$ of active and inactive edges surrounding each node, and can be straightforwardly realized as $E=|\mathcal{N}_i|$ independent binary random variables. 

However, since this sampling occurs for every neighborhood $\mathcal{N}_i$ and is repeated $M$ times, it is worthwhile to optimize.
This appendix describes practical implementations used in our simulations. 

\subsection{Breadth-first search simulations}
To sample each neighborhood, we simulate independent cascades with early stopping (also known as ``on-the-fly percolation''~\cite{noel2012propagation}), as it proves somewhat more efficient than naive percolation simulations (where the outcome of a Bernoulli trial determines the state of every edge).
The simulation begins at focal node $i$ and explores $\mathcal{N}_i$ as a breadth-first search. 
At each level of the search, branches are pruned with probability $1-p$, and pruned edges cannot be traversed in subsequent levels.
The set $N(\gamma)$ then consists of all discovered nodes,  while the level at which a node was first found corresponds to its distance $\ell$.
With these distances computed, Eq.~\eqref{eq:dynamicNMP_core} can be used directly in conjunction with the sampling approximation of Eq.~\eqref{eq:sampling_approx}.
Early stopping prevents the exploration of the whole neighborhood at low infectivity levels $p$.

\subsection{Newman-Ziff simulations}
Previous work on NMP~\cite{cantwellMessagePassingNetworks2019} makes use of the ``Newman-Ziff'' algorithm~\cite{newmanFastMonteCarlo2001} instead, as it can be faster in certain applications.
This algorithm offers a more significant speed-up when $p_{ij}=p$ for all edges, so we'll restrict ourselves to this case.
It goes as follows:
Starting with an empty graph, all edges in the neighborhood $\mathcal{N}_i$ are added in a random order (selected uniformly from all permutations). 
A union-find data structure tracks the composition of connected clusters and thus the list of nodes reachable from $i$.
Repeating this $M$ times generates $M\times E$ different percolation outcomes, where $E=|\mathcal{N}_i|$. 
Though outcomes from a single iteration are highly correlated, multiple iterations produce uncorrelated samples.

This leaves the question of the probability of the various samples.
Of the $M\times E$ outcomes, $M$ have exactly $e\in[0,E]$ edges, and the relative probability of a percolation outcome with exactly $e$ edges is
\begin{equation*}
    w_e = {E \choose e} p^e (1-p)^{(E-e)}
    \label{eq:newman_ziff_weights}
.\end{equation*}
Hence, the probability of a particular percolation outcome can be estimated as
\begin{equation*}
    P(\Gamma_i = \gamma) = \frac{w_{e(\gamma)}}{M}
,\end{equation*}
which leads to the following importance sampling approximation for NMP
\begin{equation*}
    \langle f(\Gamma_i)\rangle \approx \sum_{m=1}^M \sum_{e=0}^E f\left(\Gamma_i = \gamma_m^{(e)}\right) w_{e(\gamma)}
,\end{equation*}
where $\gamma_m^{(e)}$ is a percolation outcome with $e$ active edges from the $m$th iteration.

\begin{figure}
    \centering
    \vspace{0.2cm}
    \includegraphics[width=1.0\linewidth]{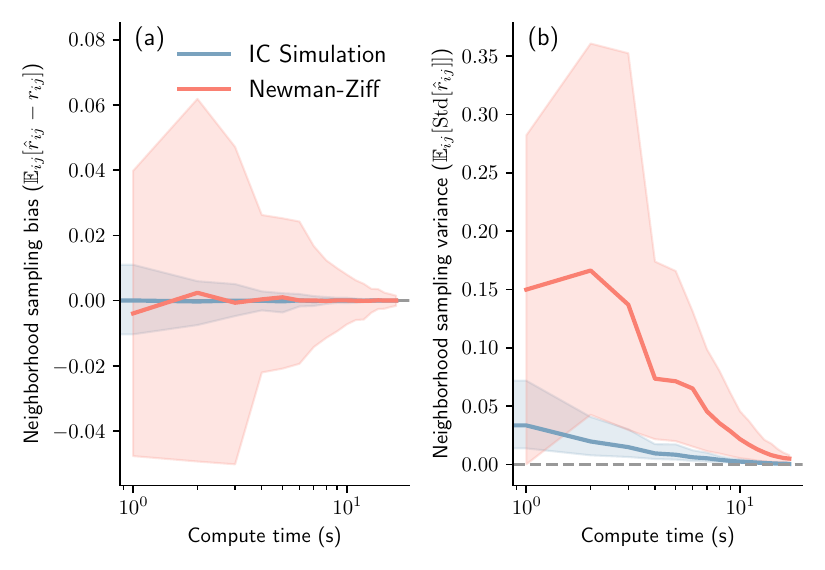}
    \caption{
    \textbf{Comparing Monte Carlo sampling algorithms.}
    We compare the performance of two sampling strategies (Newman-Ziff and IC simulation) in estimating the expected reachability $\hat{r}_{ij}$, averaging over all nodes in all neighborhoods.
    \textbf{(a)} The average bias (systematic deviation from the true value $r_{ij}$) for a random node in a random neighborhood.
    \textbf{(b)} The average standard deviation of $\hat{r}_{ij}$ for the expected reachability of a random node in a random neighborhood.
    Error bars show a 95th percentile interval.
    }
    \label{fig:sampling-bias}
\end{figure}

\subsection{Comparison with Newman-Ziff}

In this section, we compare both sampling strategies: Breadth-first search and Newman-Ziff.

We evaluate how each method estimates the probability that a node $j \in \mathcal{N}_i$ is reachable from node $i$ on a random outcome of the percolation process $\Gamma_{\mathcal{N}_i} = \gamma$, which we designate by the binary random variable $R_{ij}.$
Let $r_{ij} = \mathbb{E}_\gamma [R_{ij}]$ be the expected reachability of $j\in \mathcal{N}_i$, and let
\begin{equation*}
    \hat{r}_{ij} = \frac{1}{M} \sum_{m=1}^M \sum_{\bm{X}}\mathbb{1}\left[ j \in N_i(\gamma_m)\right]
\end{equation*}
be the sampling approximation of $r_{ij}$, computed on $M$ neighborhood samples.

In Figure \ref{fig:sampling-bias}, we compute $50$ independent sets of $M$ samples for all neighborhoods of the karate club network, varying $M$ and thus the total computation time.
Because the number of samples is not directly comparable between the Newman-Ziff and IC Monte Carlo sampling strategies, we consider only the total computation time required for each method.

We evaluated the bias of the estimator $\hat{r}_{ij}$ under each method and found that when selecting a random node in a random neighborhood, the sampling approximation of the expected reachability is unbiased for both methods.
We also find that the average standard deviation of $\hat{r}_{ij}$ is considerably larger under the Newman-Ziff algorithm---likely due to the fact that realizations within each ``sweep'' of the algorithm are highly correlated.
This result suggests that basic simulations of the independent cascade model make better use of a fixed computing budget in the context of NMP.

\begin{figure*}[b!]
    \centering
    \includegraphics[width=0.8\linewidth]{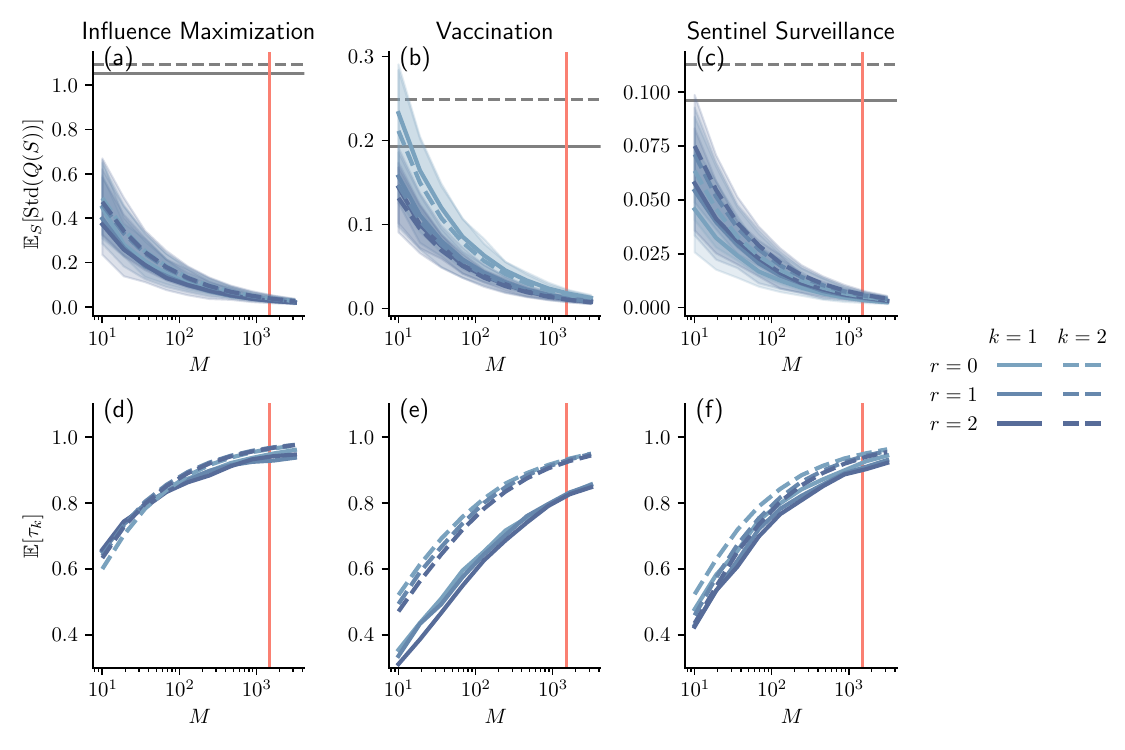}
    \caption[Comparing the required number of samples, $M$]{\textbf{Effect of the number of samples $M$ on accuracy.} 
    We run simulations on the karate club graph \cite{zachary1977_InformationFlowModel} at infection probability $p=0.15$.
    For each set $S$, we run $N=50$ instances of the NMP algorithm. 
    \textbf{(a-c)} For each set, we compute the standard deviation of the set's quality across realizations, i.e., $\mathrm{Std}(Q(S))$ and show this quantity for all sets of size $k=1$ (solid lines) and $k=2$ (dashed lines).
    The horizontal lines (solid for $k=1$ and dashed for $k=2$) show the standard deviation of the true quality scores $\mathrm{Std}_S[Q^*(S)]$ as approximated by $10^6$ Monte Carlo simulations. 
    Error bars denote the 95-percentile intervals.
    \textbf{(d-f)}
    The rank correlation of all pairs of the $50$ independent realizations of the NMP algorithm.
    The red vertical line indicates $M=1500$, the number of samples used in our experiments.
    }
    \label{fig:rT_tradeoff}
\end{figure*}

\subsection{Effect of the number of samples}
\label{appendix:costs}

As reported in \cite{cantwellMessagePassingNetworks2019}, surprisingly few Monte Carlo samples are required to obtain a good approximation of the message passing probabilities.
Figure \ref{fig:rT_tradeoff} shows that the variation in outcomes across algorithm runs declines quickly with the number of samples, though convergence is slowest for vaccination.


\begin{thebibliography}{48}%
\makeatletter
\providecommand \@ifxundefined [1]{%
 \@ifx{#1\undefined}
}%
\providecommand \@ifnum [1]{%
 \ifnum #1\expandafter \@firstoftwo
 \else \expandafter \@secondoftwo
 \fi
}%
\providecommand \@ifx [1]{%
 \ifx #1\expandafter \@firstoftwo
 \else \expandafter \@secondoftwo
 \fi
}%
\providecommand \natexlab [1]{#1}%
\providecommand \enquote  [1]{``#1''}%
\providecommand \bibnamefont  [1]{#1}%
\providecommand \bibfnamefont [1]{#1}%
\providecommand \citenamefont [1]{#1}%
\providecommand \href@noop [0]{\@secondoftwo}%
\providecommand \href [0]{\begingroup \@sanitize@url \@href}%
\providecommand \@href[1]{\@@startlink{#1}\@@href}%
\providecommand \@@href[1]{\endgroup#1\@@endlink}%
\providecommand \@sanitize@url [0]{\catcode `\\12\catcode `\$12\catcode `\&12\catcode `\#12\catcode `\^12\catcode `\_12\catcode `\%12\relax}%
\providecommand \@@startlink[1]{}%
\providecommand \@@endlink[0]{}%
\providecommand \url  [0]{\begingroup\@sanitize@url \@url }%
\providecommand \@url [1]{\endgroup\@href {#1}{\urlprefix }}%
\providecommand \urlprefix  [0]{URL }%
\providecommand \Eprint [0]{\href }%
\providecommand \doibase [0]{http://dx.doi.org/}%
\providecommand \selectlanguage [0]{\@gobble}%
\providecommand \bibinfo  [0]{\@secondoftwo}%
\providecommand \bibfield  [0]{\@secondoftwo}%
\providecommand \translation [1]{[#1]}%
\providecommand \BibitemOpen [0]{}%
\providecommand \bibitemStop [0]{}%
\providecommand \bibitemNoStop [0]{.\EOS\space}%
\providecommand \EOS [0]{\spacefactor3000\relax}%
\providecommand \BibitemShut  [1]{\csname bibitem#1\endcsname}%
\let\auto@bib@innerbib\@empty
\bibitem [{\citenamefont {Chami}\ \emph {et~al.}(2017)\citenamefont {Chami}, \citenamefont {Ahnert}, \citenamefont {Kabatereine},\ and\ \citenamefont {Tukahebwa}}]{chami2017SocialNetwork}%
  \BibitemOpen
  \bibfield  {author} {\bibinfo {author} {\bibfnamefont {G.~F.}\ \bibnamefont {Chami}}, \bibinfo {author} {\bibfnamefont {S.~E.}\ \bibnamefont {Ahnert}}, \bibinfo {author} {\bibfnamefont {N.~B.}\ \bibnamefont {Kabatereine}}, \ and\ \bibinfo {author} {\bibfnamefont {E.~M.}\ \bibnamefont {Tukahebwa}},\ }\href {\doibase 10.1073/pnas.1700166114} {\bibfield  {journal} {\bibinfo  {journal} {Proc. Natl. Acad. Sci. U.S.A.}\ }\textbf {\bibinfo {volume} {114}},\ \bibinfo {pages} {E7425} (\bibinfo {year} {2017})}\BibitemShut {NoStop}%
\bibitem [{\citenamefont {Nishi}\ \emph {et~al.}(2020)\citenamefont {Nishi}, \citenamefont {Dewey}, \citenamefont {Endo}, \citenamefont {Neman}, \citenamefont {Iwamoto}, \citenamefont {Ni}, \citenamefont {Tsugawa}, \citenamefont {Iosifidis}, \citenamefont {Smith},\ and\ \citenamefont {Young}}]{nishi2020NetworkInterventions}%
  \BibitemOpen
  \bibfield  {author} {\bibinfo {author} {\bibfnamefont {A.}~\bibnamefont {Nishi}}, \bibinfo {author} {\bibfnamefont {G.}~\bibnamefont {Dewey}}, \bibinfo {author} {\bibfnamefont {A.}~\bibnamefont {Endo}}, \bibinfo {author} {\bibfnamefont {S.}~\bibnamefont {Neman}}, \bibinfo {author} {\bibfnamefont {S.~K.}\ \bibnamefont {Iwamoto}}, \bibinfo {author} {\bibfnamefont {M.~Y.}\ \bibnamefont {Ni}}, \bibinfo {author} {\bibfnamefont {Y.}~\bibnamefont {Tsugawa}}, \bibinfo {author} {\bibfnamefont {G.}~\bibnamefont {Iosifidis}}, \bibinfo {author} {\bibfnamefont {J.~D.}\ \bibnamefont {Smith}}, \ and\ \bibinfo {author} {\bibfnamefont {S.~D.}\ \bibnamefont {Young}},\ }\href {\doibase 10.1073/pnas.2014297117} {\bibfield  {journal} {\bibinfo  {journal} {Proc. Natl. Acad. Sci. U.S.A.}\ }\textbf {\bibinfo {volume} {117}},\ \bibinfo {pages} {30285} (\bibinfo {year} {2020})}\BibitemShut {NoStop}%
\bibitem [{\citenamefont {Valente}(2012)}]{valente2012NetworkInterventions}%
  \BibitemOpen
  \bibfield  {author} {\bibinfo {author} {\bibfnamefont {T.~W.}\ \bibnamefont {Valente}},\ }\href {\doibase 10.1126/science.1217330} {\bibfield  {journal} {\bibinfo  {journal} {Science}\ }\textbf {\bibinfo {volume} {337}},\ \bibinfo {pages} {49} (\bibinfo {year} {2012})}\BibitemShut {NoStop}%
\bibitem [{\citenamefont {Aral}\ and\ \citenamefont {Dhillon}(2018{\natexlab{a}})}]{aral2018SocialInfluence}%
  \BibitemOpen
  \bibfield  {author} {\bibinfo {author} {\bibfnamefont {S.}~\bibnamefont {Aral}}\ and\ \bibinfo {author} {\bibfnamefont {P.~S.}\ \bibnamefont {Dhillon}},\ }\href {\doibase 10.1038/s41562-018-0346-z} {\bibfield  {journal} {\bibinfo  {journal} {Nat. Hum. Behav.}\ }\textbf {\bibinfo {volume} {2}},\ \bibinfo {pages} {375} (\bibinfo {year} {2018}{\natexlab{a}})}\BibitemShut {NoStop}%
\bibitem [{\citenamefont {Banerjee}\ \emph {et~al.}(2013)\citenamefont {Banerjee}, \citenamefont {Chandrasekhar}, \citenamefont {Duflo},\ and\ \citenamefont {Jackson}}]{banerjee2013DiffusionMicrofinance}%
  \BibitemOpen
  \bibfield  {author} {\bibinfo {author} {\bibfnamefont {A.}~\bibnamefont {Banerjee}}, \bibinfo {author} {\bibfnamefont {A.~G.}\ \bibnamefont {Chandrasekhar}}, \bibinfo {author} {\bibfnamefont {E.}~\bibnamefont {Duflo}}, \ and\ \bibinfo {author} {\bibfnamefont {M.~O.}\ \bibnamefont {Jackson}},\ }\href {\doibase 10.1126/science.1236498} {\bibfield  {journal} {\bibinfo  {journal} {Science}\ }\textbf {\bibinfo {volume} {341}},\ \bibinfo {pages} {1236498} (\bibinfo {year} {2013})}\BibitemShut {NoStop}%
\bibitem [{\citenamefont {Van~den Bulte}\ and\ \citenamefont {Joshi}(2007)}]{van2007ProductDiffusion}%
  \BibitemOpen
  \bibfield  {author} {\bibinfo {author} {\bibfnamefont {C.}~\bibnamefont {Van~den Bulte}}\ and\ \bibinfo {author} {\bibfnamefont {Y.~V.}\ \bibnamefont {Joshi}},\ }\href {\doibase 10.1287/mksc.1060.0224} {\bibfield  {journal} {\bibinfo  {journal} {Mark. Sci.}\ }\textbf {\bibinfo {volume} {26}},\ \bibinfo {pages} {400} (\bibinfo {year} {2007})}\BibitemShut {NoStop}%
\bibitem [{\citenamefont {Divi{\'a}k}\ \emph {et~al.}(2022)\citenamefont {Divi{\'a}k}, \citenamefont {van Nassau}, \citenamefont {Dijkstra},\ and\ \citenamefont {Snijders}}]{diviak2022DynamicsDisruption}%
  \BibitemOpen
  \bibfield  {author} {\bibinfo {author} {\bibfnamefont {T.}~\bibnamefont {Divi{\'a}k}}, \bibinfo {author} {\bibfnamefont {C.~S.}\ \bibnamefont {van Nassau}}, \bibinfo {author} {\bibfnamefont {J.~K.}\ \bibnamefont {Dijkstra}}, \ and\ \bibinfo {author} {\bibfnamefont {T.~A.}\ \bibnamefont {Snijders}},\ }\href {\doibase 10.1016/j.socnet.2022.04.001} {\bibfield  {journal} {\bibinfo  {journal} {Soc. Netw.}\ }\textbf {\bibinfo {volume} {70}},\ \bibinfo {pages} {364} (\bibinfo {year} {2022})}\BibitemShut {NoStop}%
\bibitem [{\citenamefont {Kempe}\ \emph {et~al.}(2003)\citenamefont {Kempe}, \citenamefont {Kleinberg},\ and\ \citenamefont {Tardos}}]{kempeMaximizingSpreadInfluence2003}%
  \BibitemOpen
  \bibfield  {author} {\bibinfo {author} {\bibfnamefont {D.}~\bibnamefont {Kempe}}, \bibinfo {author} {\bibfnamefont {J.}~\bibnamefont {Kleinberg}}, \ and\ \bibinfo {author} {\bibfnamefont {{\'E}.}~\bibnamefont {Tardos}},\ }in\ \href {\doibase 10.1145/956750.956769} {\emph {\bibinfo {booktitle} {Proceedings of the Ninth {{ACM SIGKDD}} International Conference on {{Knowledge}} Discovery and Data Mining}}},\ \bibinfo {series and number} {{{KDD}} '03}\ (\bibinfo  {publisher} {Association for Computing Machinery},\ \bibinfo {address} {New York, NY, USA},\ \bibinfo {year} {2003})\ pp.\ \bibinfo {pages} {137--146}\BibitemShut {NoStop}%
\bibitem [{\citenamefont {Pei}\ \emph {et~al.}(2019)\citenamefont {Pei}, \citenamefont {Wang}, \citenamefont {Morone},\ and\ \citenamefont {Makse}}]{pei2020influencer}%
  \BibitemOpen
  \bibfield  {author} {\bibinfo {author} {\bibfnamefont {S.}~\bibnamefont {Pei}}, \bibinfo {author} {\bibfnamefont {J.}~\bibnamefont {Wang}}, \bibinfo {author} {\bibfnamefont {F.}~\bibnamefont {Morone}}, \ and\ \bibinfo {author} {\bibfnamefont {H.~A.}\ \bibnamefont {Makse}},\ }\href {\doibase 10.1093/comnet/cnz029} {\bibfield  {journal} {\bibinfo  {journal} {J. Complex Netw.}\ }\textbf {\bibinfo {volume} {8}},\ \bibinfo {pages} {cnz029} (\bibinfo {year} {2019})}\BibitemShut {NoStop}%
\bibitem [{\citenamefont {Domingos}\ and\ \citenamefont {Richardson}(2001)}]{domingosMiningNetworkValue2001}%
  \BibitemOpen
  \bibfield  {author} {\bibinfo {author} {\bibfnamefont {P.}~\bibnamefont {Domingos}}\ and\ \bibinfo {author} {\bibfnamefont {M.}~\bibnamefont {Richardson}},\ }in\ \href {\doibase 10.1145/502512.502525} {\emph {\bibinfo {booktitle} {Proceedings of the Seventh {{ACM SIGKDD}} International Conference on {{Knowledge}} Discovery and Data Mining}}},\ \bibinfo {series and number} {{{KDD}} '01}\ (\bibinfo  {publisher} {{Association for Computing Machinery}},\ \bibinfo {address} {{New York, NY, USA}},\ \bibinfo {year} {2001})\ pp.\ \bibinfo {pages} {57--66}\BibitemShut {NoStop}%
\bibitem [{\citenamefont {H{\'e}bert-Dufresne}\ \emph {et~al.}(2013)\citenamefont {H{\'e}bert-Dufresne}, \citenamefont {Allard}, \citenamefont {Young},\ and\ \citenamefont {Dub{\'e}}}]{hebert2013global}%
  \BibitemOpen
  \bibfield  {author} {\bibinfo {author} {\bibfnamefont {L.}~\bibnamefont {H{\'e}bert-Dufresne}}, \bibinfo {author} {\bibfnamefont {A.}~\bibnamefont {Allard}}, \bibinfo {author} {\bibfnamefont {J.-G.}\ \bibnamefont {Young}}, \ and\ \bibinfo {author} {\bibfnamefont {L.~J.}\ \bibnamefont {Dub{\'e}}},\ }\href {\doibase 10.1038/srep02171} {\bibfield  {journal} {\bibinfo  {journal} {Sci. Rep.}\ }\textbf {\bibinfo {volume} {3}},\ \bibinfo {pages} {2171} (\bibinfo {year} {2013})}\BibitemShut {NoStop}%
\bibitem [{\citenamefont {Holme}(2017)}]{holmeThreeFacesNode2017}%
  \BibitemOpen
  \bibfield  {author} {\bibinfo {author} {\bibfnamefont {P.}~\bibnamefont {Holme}},\ }\href {\doibase 10.1103/PhysRevE.96.062305} {\bibfield  {journal} {\bibinfo  {journal} {Phys. Rev. E}\ }\textbf {\bibinfo {volume} {96}},\ \bibinfo {pages} {062305} (\bibinfo {year} {2017})}\BibitemShut {NoStop}%
\bibitem [{\citenamefont {Allen}\ \emph {et~al.}(2022)\citenamefont {Allen}, \citenamefont {Boudreau}, \citenamefont {Roberts}, \citenamefont {Allard},\ and\ \citenamefont {H{\'e}bert-Dufresne}}]{allen2022predicting}%
  \BibitemOpen
  \bibfield  {author} {\bibinfo {author} {\bibfnamefont {A.~J.}\ \bibnamefont {Allen}}, \bibinfo {author} {\bibfnamefont {M.~C.}\ \bibnamefont {Boudreau}}, \bibinfo {author} {\bibfnamefont {N.~J.}\ \bibnamefont {Roberts}}, \bibinfo {author} {\bibfnamefont {A.}~\bibnamefont {Allard}}, \ and\ \bibinfo {author} {\bibfnamefont {L.}~\bibnamefont {H{\'e}bert-Dufresne}},\ }\href {\doibase 10.1103/PhysRevResearch.4.013123} {\bibfield  {journal} {\bibinfo  {journal} {Phys. Rev. Res.}\ }\textbf {\bibinfo {volume} {4}},\ \bibinfo {pages} {013123} (\bibinfo {year} {2022})}\BibitemShut {NoStop}%
\bibitem [{\citenamefont {H{\'e}bert-Dufresne}\ \emph {et~al.}(2010)\citenamefont {H{\'e}bert-Dufresne}, \citenamefont {No{\"e}l}, \citenamefont {Marceau}, \citenamefont {Allard},\ and\ \citenamefont {Dub{\'e}}}]{hebert2010propagation}%
  \BibitemOpen
  \bibfield  {author} {\bibinfo {author} {\bibfnamefont {L.}~\bibnamefont {H{\'e}bert-Dufresne}}, \bibinfo {author} {\bibfnamefont {P.-A.}\ \bibnamefont {No{\"e}l}}, \bibinfo {author} {\bibfnamefont {V.}~\bibnamefont {Marceau}}, \bibinfo {author} {\bibfnamefont {A.}~\bibnamefont {Allard}}, \ and\ \bibinfo {author} {\bibfnamefont {L.~J.}\ \bibnamefont {Dub{\'e}}},\ }\href {\doibase 10.1103/PhysRevE.82.036115} {\bibfield  {journal} {\bibinfo  {journal} {Phys. Rev. E}\ }\textbf {\bibinfo {volume} {82}},\ \bibinfo {pages} {036115} (\bibinfo {year} {2010})}\BibitemShut {NoStop}%
\bibitem [{\citenamefont {Marceau}\ \emph {et~al.}(2010)\citenamefont {Marceau}, \citenamefont {No{\"e}l}, \citenamefont {H{\'e}bert-Dufresne}, \citenamefont {Allard},\ and\ \citenamefont {Dub{\'e}}}]{marceau2010adaptive}%
  \BibitemOpen
  \bibfield  {author} {\bibinfo {author} {\bibfnamefont {V.}~\bibnamefont {Marceau}}, \bibinfo {author} {\bibfnamefont {P.-A.}\ \bibnamefont {No{\"e}l}}, \bibinfo {author} {\bibfnamefont {L.}~\bibnamefont {H{\'e}bert-Dufresne}}, \bibinfo {author} {\bibfnamefont {A.}~\bibnamefont {Allard}}, \ and\ \bibinfo {author} {\bibfnamefont {L.~J.}\ \bibnamefont {Dub{\'e}}},\ }\href {\doibase 10.1103/PhysRevE.82.036116} {\bibfield  {journal} {\bibinfo  {journal} {Phys. Rev. E}\ }\textbf {\bibinfo {volume} {82}},\ \bibinfo {pages} {036116} (\bibinfo {year} {2010})}\BibitemShut {NoStop}%
\bibitem [{\citenamefont {St-Onge}\ \emph {et~al.}(2022)\citenamefont {St-Onge}, \citenamefont {Iacopini}, \citenamefont {Latora}, \citenamefont {Barrat}, \citenamefont {Petri}, \citenamefont {Allard},\ and\ \citenamefont {H{\'e}bert-Dufresne}}]{stonge2022influential}%
  \BibitemOpen
  \bibfield  {author} {\bibinfo {author} {\bibfnamefont {G.}~\bibnamefont {St-Onge}}, \bibinfo {author} {\bibfnamefont {I.}~\bibnamefont {Iacopini}}, \bibinfo {author} {\bibfnamefont {V.}~\bibnamefont {Latora}}, \bibinfo {author} {\bibfnamefont {A.}~\bibnamefont {Barrat}}, \bibinfo {author} {\bibfnamefont {G.}~\bibnamefont {Petri}}, \bibinfo {author} {\bibfnamefont {A.}~\bibnamefont {Allard}}, \ and\ \bibinfo {author} {\bibfnamefont {L.}~\bibnamefont {H{\'e}bert-Dufresne}},\ }\href@noop {} {\bibfield  {journal} {\bibinfo  {journal} {Communications Physics}\ }\textbf {\bibinfo {volume} {5}},\ \bibinfo {pages} {25} (\bibinfo {year} {2022})}\BibitemShut {NoStop}%
\bibitem [{\citenamefont {Shrestha}\ \emph {et~al.}(2015)\citenamefont {Shrestha}, \citenamefont {Scarpino},\ and\ \citenamefont {Moore}}]{shresthaMessagepassingApproachRecurrentstate2015}%
  \BibitemOpen
  \bibfield  {author} {\bibinfo {author} {\bibfnamefont {M.}~\bibnamefont {Shrestha}}, \bibinfo {author} {\bibfnamefont {S.~V.}\ \bibnamefont {Scarpino}}, \ and\ \bibinfo {author} {\bibfnamefont {C.}~\bibnamefont {Moore}},\ }\href {\doibase 10.1103/PhysRevE.92.022821} {\bibfield  {journal} {\bibinfo  {journal} {Phys. Rev. E}\ }\textbf {\bibinfo {volume} {92}},\ \bibinfo {pages} {022821} (\bibinfo {year} {2015})}\BibitemShut {NoStop}%
\bibitem [{\citenamefont {Pearl}(1982)}]{pearlReverendBayesInference1982}%
  \BibitemOpen
  \bibfield  {author} {\bibinfo {author} {\bibfnamefont {J.}~\bibnamefont {Pearl}},\ }in\ \href {https://dl.acm.org/doi/10.5555/2876686.2876719} {\emph {\bibinfo {booktitle} {Proceedings of the {{Second AAAI Conference}} on {{Artificial Intelligence}}}}},\ \bibinfo {series and number} {{{AAAI}}'82}\ (\bibinfo  {publisher} {AAAI Press},\ \bibinfo {address} {Pittsburgh, Pennsylvania},\ \bibinfo {year} {1982})\ pp.\ \bibinfo {pages} {133--136}\BibitemShut {NoStop}%
\bibitem [{\citenamefont {Mézard}\ and\ \citenamefont {Montanari}(2009)}]{mezardInformationPhysicsComputation2009a}%
  \BibitemOpen
  \bibfield  {author} {\bibinfo {author} {\bibfnamefont {M.}~\bibnamefont {Mézard}}\ and\ \bibinfo {author} {\bibfnamefont {A.}~\bibnamefont {Montanari}},\ }\href@noop {} {\emph {\bibinfo {title} {Information, {{Physics}}, and {{Computation}}}}}\ (\bibinfo  {publisher} {Oxford University Press},\ \bibinfo {address} {New York, NY, USA},\ \bibinfo {year} {2009})\BibitemShut {NoStop}%
\bibitem [{\citenamefont {Karrer}\ and\ \citenamefont {Newman}(2010)}]{karrerMessagePassingApproach2010}%
  \BibitemOpen
  \bibfield  {author} {\bibinfo {author} {\bibfnamefont {B.}~\bibnamefont {Karrer}}\ and\ \bibinfo {author} {\bibfnamefont {M.~E.~J.}\ \bibnamefont {Newman}},\ }\href {\doibase 10.1103/PhysRevE.82.016101} {\bibfield  {journal} {\bibinfo  {journal} {Phys. Rev. E}\ }\textbf {\bibinfo {volume} {82}},\ \bibinfo {pages} {016101} (\bibinfo {year} {2010})}\BibitemShut {NoStop}%
\bibitem [{\citenamefont {Lokhov}\ \emph {et~al.}(2014)\citenamefont {Lokhov}, \citenamefont {M{\'e}zard}, \citenamefont {Ohta},\ and\ \citenamefont {Zdeborov{\'a}}}]{lokhovInferringOriginEpidemic2014}%
  \BibitemOpen
  \bibfield  {author} {\bibinfo {author} {\bibfnamefont {A.~Y.}\ \bibnamefont {Lokhov}}, \bibinfo {author} {\bibfnamefont {M.}~\bibnamefont {M{\'e}zard}}, \bibinfo {author} {\bibfnamefont {H.}~\bibnamefont {Ohta}}, \ and\ \bibinfo {author} {\bibfnamefont {L.}~\bibnamefont {Zdeborov{\'a}}},\ }\href {\doibase 10.1103/PhysRevE.90.012801} {\bibfield  {journal} {\bibinfo  {journal} {Phys. Rev. E}\ }\textbf {\bibinfo {volume} {90}},\ \bibinfo {pages} {012801} (\bibinfo {year} {2014})}\BibitemShut {NoStop}%
\bibitem [{\citenamefont {Newman}(2023)}]{newmanMessagePassingMethods2023}%
  \BibitemOpen
  \bibfield  {author} {\bibinfo {author} {\bibfnamefont {M.~E.~J.}\ \bibnamefont {Newman}},\ }\href {\doibase 10.1098/rspa.2022.0774} {\bibfield  {journal} {\bibinfo  {journal} {Proc. R. Soc. A}\ }\textbf {\bibinfo {volume} {479}},\ \bibinfo {pages} {20220774} (\bibinfo {year} {2023})}\BibitemShut {NoStop}%
\bibitem [{\citenamefont {Altarelli}\ \emph {et~al.}(2014)\citenamefont {Altarelli}, \citenamefont {Braunstein}, \citenamefont {Dall'Asta}, \citenamefont {Wakeling},\ and\ \citenamefont {Zecchina}}]{altarelliContainingEpidemicOutbreaks2014}%
  \BibitemOpen
  \bibfield  {author} {\bibinfo {author} {\bibfnamefont {F.}~\bibnamefont {Altarelli}}, \bibinfo {author} {\bibfnamefont {A.}~\bibnamefont {Braunstein}}, \bibinfo {author} {\bibfnamefont {L.}~\bibnamefont {Dall'Asta}}, \bibinfo {author} {\bibfnamefont {J.~R.}\ \bibnamefont {Wakeling}}, \ and\ \bibinfo {author} {\bibfnamefont {R.}~\bibnamefont {Zecchina}},\ }\href {\doibase 10.1103/PhysRevX.4.021024} {\bibfield  {journal} {\bibinfo  {journal} {Phys. Rev. X}\ }\textbf {\bibinfo {volume} {4}},\ \bibinfo {pages} {021024} (\bibinfo {year} {2014})},\ \Eprint {http://arxiv.org/abs/1309.2805} {1309.2805} \BibitemShut {NoStop}%
\bibitem [{\citenamefont {Morone}\ and\ \citenamefont {Makse}(2015)}]{moroneInfluenceMaximizationComplex2015}%
  \BibitemOpen
  \bibfield  {author} {\bibinfo {author} {\bibfnamefont {F.}~\bibnamefont {Morone}}\ and\ \bibinfo {author} {\bibfnamefont {H.~A.}\ \bibnamefont {Makse}},\ }\href {\doibase 10.1038/nature14604} {\bibfield  {journal} {\bibinfo  {journal} {Nature}\ }\textbf {\bibinfo {volume} {524}},\ \bibinfo {pages} {65} (\bibinfo {year} {2015})}\BibitemShut {NoStop}%
\bibitem [{\citenamefont {Karrer}\ \emph {et~al.}(2014)\citenamefont {Karrer}, \citenamefont {Newman},\ and\ \citenamefont {Zdeborov{\'a}}}]{karrerPercolationSparseNetworks2014}%
  \BibitemOpen
  \bibfield  {author} {\bibinfo {author} {\bibfnamefont {B.}~\bibnamefont {Karrer}}, \bibinfo {author} {\bibfnamefont {M.~E.~J.}\ \bibnamefont {Newman}}, \ and\ \bibinfo {author} {\bibfnamefont {L.}~\bibnamefont {Zdeborov{\'a}}},\ }\href {\doibase 10.1103/PhysRevLett.113.208702} {\bibfield  {journal} {\bibinfo  {journal} {Phys. Rev. Lett.}\ }\textbf {\bibinfo {volume} {113}},\ \bibinfo {pages} {208702} (\bibinfo {year} {2014})}\BibitemShut {NoStop}%
\bibitem [{\citenamefont {Decelle}\ \emph {et~al.}(2011)\citenamefont {Decelle}, \citenamefont {Krzakala}, \citenamefont {Moore},\ and\ \citenamefont {Zdeborov{\'a}}}]{decelleInferencePhaseTransitions2011}%
  \BibitemOpen
  \bibfield  {author} {\bibinfo {author} {\bibfnamefont {A.}~\bibnamefont {Decelle}}, \bibinfo {author} {\bibfnamefont {F.}~\bibnamefont {Krzakala}}, \bibinfo {author} {\bibfnamefont {C.}~\bibnamefont {Moore}}, \ and\ \bibinfo {author} {\bibfnamefont {L.}~\bibnamefont {Zdeborov{\'a}}},\ }\href {\doibase 10.1103/PhysRevLett.107.065701} {\bibfield  {journal} {\bibinfo  {journal} {Phys. Rev. Lett.}\ }\textbf {\bibinfo {volume} {107}},\ \bibinfo {pages} {065701} (\bibinfo {year} {2011})}\BibitemShut {NoStop}%
\bibitem [{\citenamefont {Newman}(2018)}]{newman2018networks}%
  \BibitemOpen
  \bibfield  {author} {\bibinfo {author} {\bibfnamefont {M.}~\bibnamefont {Newman}},\ }\href@noop {} {\emph {\bibinfo {title} {Networks}}},\ \bibinfo {edition} {2nd}\ ed.\ (\bibinfo  {publisher} {Oxford University Press},\ \bibinfo {address} {New York, NY, USA},\ \bibinfo {year} {2018})\BibitemShut {NoStop}%
\bibitem [{\citenamefont {Cantwell}\ and\ \citenamefont {Newman}(2019)}]{cantwellMessagePassingNetworks2019}%
  \BibitemOpen
  \bibfield  {author} {\bibinfo {author} {\bibfnamefont {G.~T.}\ \bibnamefont {Cantwell}}\ and\ \bibinfo {author} {\bibfnamefont {M.~E.~J.}\ \bibnamefont {Newman}},\ }\href {\doibase 10.1073/pnas.1914893116} {\bibfield  {journal} {\bibinfo  {journal} {Proc. Natl. Acad. Sci. U.S.A.}\ }\textbf {\bibinfo {volume} {116}},\ \bibinfo {pages} {23398} (\bibinfo {year} {2019})}\BibitemShut {NoStop}%
\bibitem [{\citenamefont {Kirkley}\ \emph {et~al.}(2021)\citenamefont {Kirkley}, \citenamefont {Cantwell},\ and\ \citenamefont {Newman}}]{kirkleyBeliefPropagationNetworks2021}%
  \BibitemOpen
  \bibfield  {author} {\bibinfo {author} {\bibfnamefont {A.}~\bibnamefont {Kirkley}}, \bibinfo {author} {\bibfnamefont {G.~T.}\ \bibnamefont {Cantwell}}, \ and\ \bibinfo {author} {\bibfnamefont {M.~E.~J.}\ \bibnamefont {Newman}},\ }\href {\doibase 10.1126/sciadv.abf1211} {\bibfield  {journal} {\bibinfo  {journal} {Sci. Adv.}\ }\textbf {\bibinfo {volume} {7}},\ \bibinfo {pages} {eabf1211} (\bibinfo {year} {2021})}\BibitemShut {NoStop}%
\bibitem [{\citenamefont {Cantwell}\ \emph {et~al.}(2023)\citenamefont {Cantwell}, \citenamefont {Kirkley},\ and\ \citenamefont {Radicchi}}]{cantwell2023heterogeneous}%
  \BibitemOpen
  \bibfield  {author} {\bibinfo {author} {\bibfnamefont {G.~T.}\ \bibnamefont {Cantwell}}, \bibinfo {author} {\bibfnamefont {A.}~\bibnamefont {Kirkley}}, \ and\ \bibinfo {author} {\bibfnamefont {F.}~\bibnamefont {Radicchi}},\ }\href {\doibase 10.1103/PhysRevE.108.034310} {\bibfield  {journal} {\bibinfo  {journal} {Phys. Rev. E}\ }\textbf {\bibinfo {volume} {108}},\ \bibinfo {pages} {034310} (\bibinfo {year} {2023})}\BibitemShut {NoStop}%
\bibitem [{\citenamefont {Aral}\ and\ \citenamefont {Dhillon}(2018{\natexlab{b}})}]{aral2018social}%
  \BibitemOpen
  \bibfield  {author} {\bibinfo {author} {\bibfnamefont {S.}~\bibnamefont {Aral}}\ and\ \bibinfo {author} {\bibfnamefont {P.~S.}\ \bibnamefont {Dhillon}},\ }\href {\doibase 10.1038/s41562-018-0346-z} {\bibfield  {journal} {\bibinfo  {journal} {Nat. Hum. Behav.}\ }\textbf {\bibinfo {volume} {2}},\ \bibinfo {pages} {375} (\bibinfo {year} {2018}{\natexlab{b}})}\BibitemShut {NoStop}%
\bibitem [{\citenamefont {Newman}(2002)}]{newmanSpreadEpidemicDisease2002}%
  \BibitemOpen
  \bibfield  {author} {\bibinfo {author} {\bibfnamefont {M.~E.~J.}\ \bibnamefont {Newman}},\ }\href {\doibase 10.1103/PhysRevE.66.016128} {\bibfield  {journal} {\bibinfo  {journal} {Phys. Rev. E}\ }\textbf {\bibinfo {volume} {66}},\ \bibinfo {pages} {016128} (\bibinfo {year} {2002})}\BibitemShut {NoStop}%
\bibitem [{\citenamefont {Kenah}\ and\ \citenamefont {Robins}(2007)}]{kenahSecondLookSpread2007b}%
  \BibitemOpen
  \bibfield  {author} {\bibinfo {author} {\bibfnamefont {E.}~\bibnamefont {Kenah}}\ and\ \bibinfo {author} {\bibfnamefont {J.~M.}\ \bibnamefont {Robins}},\ }\href {\doibase 10.1103/PhysRevE.76.036113} {\bibfield  {journal} {\bibinfo  {journal} {Phys. Rev. E}\ }\textbf {\bibinfo {volume} {76}},\ \bibinfo {pages} {036113} (\bibinfo {year} {2007})}\BibitemShut {NoStop}%
\bibitem [{\citenamefont {Melnik}\ \emph {et~al.}(2011)\citenamefont {Melnik}, \citenamefont {Hackett}, \citenamefont {Porter}, \citenamefont {Mucha},\ and\ \citenamefont {Gleeson}}]{melnikUnreasonableEffectivenessTreebased2011}%
  \BibitemOpen
  \bibfield  {author} {\bibinfo {author} {\bibfnamefont {S.}~\bibnamefont {Melnik}}, \bibinfo {author} {\bibfnamefont {A.}~\bibnamefont {Hackett}}, \bibinfo {author} {\bibfnamefont {M.~A.}\ \bibnamefont {Porter}}, \bibinfo {author} {\bibfnamefont {P.~J.}\ \bibnamefont {Mucha}}, \ and\ \bibinfo {author} {\bibfnamefont {J.~P.}\ \bibnamefont {Gleeson}},\ }\href {\doibase 10.1103/PhysRevE.83.036112} {\bibfield  {journal} {\bibinfo  {journal} {Phys. Rev. E}\ }\textbf {\bibinfo {volume} {83}},\ \bibinfo {pages} {036112} (\bibinfo {year} {2011})}\BibitemShut {NoStop}%
\bibitem [{\citenamefont {Allard}\ and\ \citenamefont {H{\'e}bert-Dufresne}(2019)}]{allardAccuracyMessagepassingApproaches2019}%
  \BibitemOpen
  \bibfield  {author} {\bibinfo {author} {\bibfnamefont {A.}~\bibnamefont {Allard}}\ and\ \bibinfo {author} {\bibfnamefont {L.}~\bibnamefont {H{\'e}bert-Dufresne}},\ }\href {https://doi.org/10.48550/arXiv.1906.10377} {\bibfield  {journal} {\bibinfo  {journal} {arXiv:1906.10377}\ } (\bibinfo {year} {2019})}\BibitemShut {NoStop}%
\bibitem [{\citenamefont {Allard}\ \emph {et~al.}(2015)\citenamefont {Allard}, \citenamefont {H{\'e}bert-Dufresne}, \citenamefont {Young},\ and\ \citenamefont {Dub{\'e}}}]{allardGeneralExact2015}%
  \BibitemOpen
  \bibfield  {author} {\bibinfo {author} {\bibfnamefont {A.}~\bibnamefont {Allard}}, \bibinfo {author} {\bibfnamefont {L.}~\bibnamefont {H{\'e}bert-Dufresne}}, \bibinfo {author} {\bibfnamefont {J.-G.}\ \bibnamefont {Young}}, \ and\ \bibinfo {author} {\bibfnamefont {L.~J.}\ \bibnamefont {Dub{\'e}}},\ }\href {\doibase 10.1103/PhysRevE.92.062807} {\bibfield  {journal} {\bibinfo  {journal} {Phys. Rev. E}\ }\textbf {\bibinfo {volume} {92}},\ \bibinfo {pages} {062807} (\bibinfo {year} {2015})}\BibitemShut {NoStop}%
\bibitem [{\citenamefont {Allard}\ \emph {et~al.}(2009)\citenamefont {Allard}, \citenamefont {No{\"e}l}, \citenamefont {Dub{\'e}},\ and\ \citenamefont {Pourbohloul}}]{allard2009heterogeneous}%
  \BibitemOpen
  \bibfield  {author} {\bibinfo {author} {\bibfnamefont {A.}~\bibnamefont {Allard}}, \bibinfo {author} {\bibfnamefont {P.-A.}\ \bibnamefont {No{\"e}l}}, \bibinfo {author} {\bibfnamefont {L.~J.}\ \bibnamefont {Dub{\'e}}}, \ and\ \bibinfo {author} {\bibfnamefont {B.}~\bibnamefont {Pourbohloul}},\ }\href {\doibase 10.1103/PhysRevE.79.036113} {\bibfield  {journal} {\bibinfo  {journal} {Phys. Rev. E}\ }\textbf {\bibinfo {volume} {79}},\ \bibinfo {pages} {036113} (\bibinfo {year} {2009})}\BibitemShut {NoStop}%
\bibitem [{\citenamefont {Weis}(2025)}]{zenodo}%
  \BibitemOpen
  \bibfield  {author} {\bibinfo {author} {\bibfnamefont {E.}~\bibnamefont {Weis}},\ }\href {\doibase 10.5281/zenodo.14722576} {\enquote {\bibinfo {title} {Message passing for epidemiological interventions on networks with loops},}\ } (\bibinfo {year} {2025})\BibitemShut {NoStop}%
\bibitem [{\citenamefont {Newman}\ and\ \citenamefont {Ziff}(2001)}]{newmanFastMonteCarlo2001}%
  \BibitemOpen
  \bibfield  {author} {\bibinfo {author} {\bibfnamefont {M.~E.~J.}\ \bibnamefont {Newman}}\ and\ \bibinfo {author} {\bibfnamefont {R.~M.}\ \bibnamefont {Ziff}},\ }\href {\doibase 10.1103/PhysRevE.64.016706} {\bibfield  {journal} {\bibinfo  {journal} {Phys. Rev. E}\ }\textbf {\bibinfo {volume} {64}},\ \bibinfo {pages} {016706} (\bibinfo {year} {2001})}\BibitemShut {NoStop}%
\bibitem [{\citenamefont {Weis}(2024)}]{weis2024robust}%
  \BibitemOpen
  \bibfield  {author} {\bibinfo {author} {\bibfnamefont {E.}~\bibnamefont {Weis}},\ }\emph {\bibinfo {title} {{Robust Interventions in Network Epidemiology}}},\ \href {https://scholarworks.uvm.edu/graddis/1813/} {\bibinfo {type} {{M.Sc. Thesis}}},\ \bibinfo  {school} {University of Vermont} (\bibinfo {year} {2024})\BibitemShut {NoStop}%
\bibitem [{\citenamefont {Zachary}(1977)}]{zachary1977_InformationFlowModel}%
  \BibitemOpen
  \bibfield  {author} {\bibinfo {author} {\bibfnamefont {W.~W.}\ \bibnamefont {Zachary}},\ }\href {\doibase 10.1086/jar.33.4.3629752} {\bibfield  {journal} {\bibinfo  {journal} {J. Anthropol. Res.}\ }\textbf {\bibinfo {volume} {33}},\ \bibinfo {pages} {452} (\bibinfo {year} {1977})}\BibitemShut {NoStop}%
\bibitem [{\citenamefont {Khim}\ \emph {et~al.}(2019)\citenamefont {Khim}, \citenamefont {Jog},\ and\ \citenamefont {Loh}}]{khim2019adversarial}%
  \BibitemOpen
  \bibfield  {author} {\bibinfo {author} {\bibfnamefont {J.}~\bibnamefont {Khim}}, \bibinfo {author} {\bibfnamefont {V.}~\bibnamefont {Jog}}, \ and\ \bibinfo {author} {\bibfnamefont {P.-L.}\ \bibnamefont {Loh}},\ }in\ \href@noop {} {\emph {\bibinfo {booktitle} {{2019 IEEE International Symposium on Information Theory (ISIT)}}}}\ (\bibinfo {organization} {IEEE},\ \bibinfo {year} {2019})\ pp.\ \bibinfo {pages} {1--5}\BibitemShut {NoStop}%
\bibitem [{\citenamefont {Gozzi}\ \emph {et~al.}(2023)\citenamefont {Gozzi}, \citenamefont {Chinazzi}, \citenamefont {Dean}, \citenamefont {Longini~Jr}, \citenamefont {Halloran}, \citenamefont {Perra},\ and\ \citenamefont {Vespignani}}]{gozzi2023estimating}%
  \BibitemOpen
  \bibfield  {author} {\bibinfo {author} {\bibfnamefont {N.}~\bibnamefont {Gozzi}}, \bibinfo {author} {\bibfnamefont {M.}~\bibnamefont {Chinazzi}}, \bibinfo {author} {\bibfnamefont {N.~E.}\ \bibnamefont {Dean}}, \bibinfo {author} {\bibfnamefont {I.~M.}\ \bibnamefont {Longini~Jr}}, \bibinfo {author} {\bibfnamefont {M.~E.}\ \bibnamefont {Halloran}}, \bibinfo {author} {\bibfnamefont {N.}~\bibnamefont {Perra}}, \ and\ \bibinfo {author} {\bibfnamefont {A.}~\bibnamefont {Vespignani}},\ }\href {\doibase 10.1038/s41467-023-39098-w} {\bibfield  {journal} {\bibinfo  {journal} {Nat. Commun.}\ }\textbf {\bibinfo {volume} {14}},\ \bibinfo {pages} {3272} (\bibinfo {year} {2023})}\BibitemShut {NoStop}%
\bibitem [{\citenamefont {Radicchi}\ and\ \citenamefont {Castellano}(2016)}]{radicchiLocallyTreelikeApproximation2016}%
  \BibitemOpen
  \bibfield  {author} {\bibinfo {author} {\bibfnamefont {F.}~\bibnamefont {Radicchi}}\ and\ \bibinfo {author} {\bibfnamefont {C.}~\bibnamefont {Castellano}},\ }\href {\doibase 10.1103/PhysRevE.93.030302} {\bibfield  {journal} {\bibinfo  {journal} {Phys. Rev. E}\ }\textbf {\bibinfo {volume} {93}},\ \bibinfo {pages} {030302} (\bibinfo {year} {2016})}\BibitemShut {NoStop}%
\bibitem [{\citenamefont {Yoon}\ \emph {et~al.}(2011)\citenamefont {Yoon}, \citenamefont {Goltsev}, \citenamefont {Dorogovtsev},\ and\ \citenamefont {Mendes}}]{yoonBeliefpropagation2011}%
  \BibitemOpen
  \bibfield  {author} {\bibinfo {author} {\bibfnamefont {S.}~\bibnamefont {Yoon}}, \bibinfo {author} {\bibfnamefont {A.~V.}\ \bibnamefont {Goltsev}}, \bibinfo {author} {\bibfnamefont {S.~N.}\ \bibnamefont {Dorogovtsev}}, \ and\ \bibinfo {author} {\bibfnamefont {J.}~\bibnamefont {Mendes}},\ }\href {\doibase 10.1103/PhysRevE.84.041144} {\bibfield  {journal} {\bibinfo  {journal} {Phys. Rev. E}\ }\textbf {\bibinfo {volume} {84}},\ \bibinfo {pages} {041144} (\bibinfo {year} {2011})}\BibitemShut {NoStop}%
\bibitem [{\citenamefont {Laurence}\ \emph {et~al.}(2018)\citenamefont {Laurence}, \citenamefont {Young}, \citenamefont {Melnik},\ and\ \citenamefont {Dub{\'e}}}]{laurence2018exact}%
  \BibitemOpen
  \bibfield  {author} {\bibinfo {author} {\bibfnamefont {E.}~\bibnamefont {Laurence}}, \bibinfo {author} {\bibfnamefont {J.-G.}\ \bibnamefont {Young}}, \bibinfo {author} {\bibfnamefont {S.}~\bibnamefont {Melnik}}, \ and\ \bibinfo {author} {\bibfnamefont {L.~J.}\ \bibnamefont {Dub{\'e}}},\ }\href {\doibase PhysRevE.97.032302} {\bibfield  {journal} {\bibinfo  {journal} {Phys. Rev. E}\ }\textbf {\bibinfo {volume} {97}},\ \bibinfo {pages} {032302} (\bibinfo {year} {2018})}\BibitemShut {NoStop}%
\bibitem [{\citenamefont {Lokhov}\ and\ \citenamefont {Saad}(2017)}]{lokhov2017optimal}%
  \BibitemOpen
  \bibfield  {author} {\bibinfo {author} {\bibfnamefont {A.~Y.}\ \bibnamefont {Lokhov}}\ and\ \bibinfo {author} {\bibfnamefont {D.}~\bibnamefont {Saad}},\ }\href {\doibase 10.1073/pnas.1614694114} {\bibfield  {journal} {\bibinfo  {journal} {Proc. Natl. Acad. Sci. U.S.A.}\ }\textbf {\bibinfo {volume} {114}},\ \bibinfo {pages} {E8138} (\bibinfo {year} {2017})}\BibitemShut {NoStop}%
\bibitem [{\citenamefont {No{\"e}l}\ \emph {et~al.}(2012)\citenamefont {No{\"e}l}, \citenamefont {Allard}, \citenamefont {H{\'e}bert-Dufresne}, \citenamefont {Marceau},\ and\ \citenamefont {Dub{\'e}}}]{noel2012propagation}%
  \BibitemOpen
  \bibfield  {author} {\bibinfo {author} {\bibfnamefont {P.-A.}\ \bibnamefont {No{\"e}l}}, \bibinfo {author} {\bibfnamefont {A.}~\bibnamefont {Allard}}, \bibinfo {author} {\bibfnamefont {L.}~\bibnamefont {H{\'e}bert-Dufresne}}, \bibinfo {author} {\bibfnamefont {V.}~\bibnamefont {Marceau}}, \ and\ \bibinfo {author} {\bibfnamefont {L.~J.}\ \bibnamefont {Dub{\'e}}},\ }\href {\doibase 10.1103/PhysRevE.85.031118} {\bibfield  {journal} {\bibinfo  {journal} {Phys. Rev. E}\ }\textbf {\bibinfo {volume} {85}},\ \bibinfo {pages} {031118} (\bibinfo {year} {2012})}\BibitemShut {NoStop}%
\end{thebibliography}
\end{document}